\documentclass[acmsmall]{acmart}

\acmJournal{PACMPL}
\acmVolume{1}
\acmNumber{CONF} 
\acmArticle{1}
\acmYear{2018}
\acmMonth{1}
\acmDOI{} 
\startPage{1}

\setcopyright{none}

\usepackage{booktabs}   
\usepackage{subcaption} 
\usepackage{okexplain}

\allowdisplaybreaks
\usepackage{xcolor}
\usepackage{amsmath,amsthm}
\usepackage{mathtools}
\usepackage{suffix}
\usepackage{amsfonts}
\usepackage{mathpartir}
\usepackage{enumitem}
\usepackage{stmaryrd}
\usepackage[all]{xy}

\usepackage{amssymb}
\usepackage{twoopt}
\usepackage{tikz-cd}
\usepackage{amsthm}
\usepackage{okcategory}

\newtheorem*{remark*}{Remark}
\newtheorem*{theorem*}{Theorem}
\newtheorem*{definition*}{Definition}

\tikzset{pullback/.style={minimum size=1.2ex,path picture={
\draw[opacity=1,black,-,#1] (-0.5ex,-0.5ex) -- (0.5ex,-0.5ex) -- (0.5ex,0.5ex);%
}}}

\newcommand\insertdiagram[1]{\begin{center}
  \insertdiagram*{#1}
\end{center}}
\WithSuffix\newcommand\insertdiagram*[1]{\begin{tabular}[c]{@{}c@{}}\includegraphics{diagram-#1.mps}\end{tabular}}

\makeatletter
\newcommand\@TyAlph[1]{%
\ifcase #1\or \tau\or \sigma\or \rho\else \@ctrerr \fi%
}
\newcommand\ty[1][1]{{\@TyAlph{#1}}}

\newcommand\tvar[1][1]{{\@TyVarAlph{#1}}}
\newcommand\@TyVarAlph[1]{%
\ifcase #1\or \alpha\or \beta\or \gamma\else \@ctrerr \fi%
}

\newcommand\var[1][1]{{\@VarAlph{#1}}}
\newcommand\@VarAlph[1]{%
\ifcase #1\or x\or y\or z\or u\or v\or w\else \@ctrerr \fi%
}

\newcommand\trm[1][1]{{\@TermAlph{#1}}}
\newcommand\@TermAlph[1]{%
\ifcase #1\or t\or s\or r\else \@ctrerr \fi%
}

\newcommand\val[1][1]{%
\ifcase #1\or v\or w\or u\else \@ctrerr \fi%
}

\newcommand\op[1][1]{%
\ifcase #1\or \mathrm{op}\or \mathrm{op}'\or \mathrm{op}''\else \@ctrerr \fi%
}
\makeatother

\newcommand\Op{\mathrm{Op}}

\newcommand\sigmoid{\varsigma}

\newcommand\tUnit{\tTuple{\,}}
\newcommand\tPair[2]{\langle #1, #2\rangle}

\newcommand\tTuple[1]{\langle #1\rangle}

\newcommand\fun[1]{\lambda #1.}
\newcommand\toin[3]{#2\,\mathbf{to}\,#1.\,#3}
\newcommand\letin[3]{\mathbf{let}\,#1=\,#2\,\mathbf{in}\,#3}

\newcommand\ifelse[3]{\mathbf{if}\,#1\,\mathbf{then}\,#2\,\mathbf{else}\,#3\,}
\newcommand\rec[2]{\mu#1.#2}
\newcommand\tRoll[2][]{#1\mathbf{roll}\,#2}
\newcommand\tReturn[1]{\mathbf{return}\,#1}

\newcommand\tItFrom[3]{\mathbf{iterate}\,#1\,\mathbf{from}\,#2=#3}
\newcommand\tSign{\mathbf{sign}\,}

\newcommand\uMatch[3][\,]{\mathbf{case}\,#2\,\mathbf{of}#1\tUnit\To#3}
\newcommand\pMatch[5][\,]{\mathbf{case}\,#2\,\mathbf{of}#1\tPair{#3}{#4}\To#5}

\newcommand\vMatch[3][\,]{\mathbf{case}\,#2\,\mathbf{of}#1\{#3\}}
\newcommand\nvMatch[2][\,]{\mathbf{case}\,#2\,\mathbf{of}#1\{\,\}}
\newcommand\bvMatch[6][\,]{\mathbf{case}\,#2\,\mathbf{of}#1\{\tInl#3\To #4\, \mid \tInr#5\To #6\}}
\newcommand\rMatch[4][\,]{\mathbf{case}\,#2\,\mathbf{of}#1\mathbf{roll}\,#3\To#4}

\newcommand\tInl{\mathbf{inl}\,}
\newcommand\tInr{\mathbf{inr}\,}

\newcommand\ctx{\Gamma}
\newcommand\tinf{\vdash}
\newcommand\Ginfv[3][]{\ctx #1\tinf^v #2 : #3}
\newcommand\Ginfc[3][]{\ctx #1\tinf^c #2 : #3}
\newcommand\subst[2]{#1{}[#2]}
\newcommand\sfor[2]{^{#2}\!/\!_{#1}}

\newcommand\nat{\mathbf{nat}}
\renewcommand\reals{\mathbf{real}}

\newcommand\Unit{\mathbf{1}}

\newcommand\Init{\mathbf{0}}
\WithSuffix\newcommand\t+{\boldsymbol{\mathop{+}}}
\WithSuffix\newcommand\t*{\boldsymbol{\mathop{*}}}

\newcommand\To{\to}

\newcommand\Dsynsymbol[1][]{\scalebox{0.8}{$\overrightarrow{\mathcal{D}}$}_{#1}}
\newcommand\Dsyn[2][]{\Dsynsymbol[#1](#2)}

\newcommand\DsynV[2][]{\Dsynsymbol[#1]{}_{\Cat{V}}(#2)}
\newcommand\DsynC[2][]{\Dsynsymbol[#1]{}_{\Cat{C}}(#2)}

\newcommand\Syn{\mathbf{Syn}}
\newcommand\SynV{\Syn_V}
\newcommand\SynC{\Syn_C}

\newcommand\tFromMaybe[1]{\mathrm{fromMaybe}}

\newcommand\tMap[2]{\mathrm{map}}

\newcommand{\plots}[1]{\mathcal{P}_{#1}}

\newcommand\freeeq[1]{\stackrel{\# #1}{=}}
\newcommand\beeq{\stackrel{\beta\eta}{=}}

\definecolor{shade}{RGB}{223,223,223}
\definecolor{unshade}{RGB}{255,255,255}

\usepackage{tcolorbox}
\newtcbox{\shadebox}{on line,arc=1pt, outer arc=2pt,%
  colback=shade,colframe=shade,boxsep=0pt,%
  left=1pt,right=1pt,top=2pt,bottom=2pt,%
  boxrule=0pt,bottomrule=1pt,toprule=1pt}

\newtcbox{\unshadebox}{on line,arc=1pt, outer arc=2pt,%
  colback=unshade,colframe=shade,boxsep=0pt,%
  left=1pt,right=1pt,top=2pt,bottom=2pt,%
  boxrule=0pt,bottomrule=1pt,toprule=1pt}

\newcommand\syncat[1]{\mspace{-25mu}\synname{#1}}
\newcommand\synname[1]{\qquad\text{#1}}

\newenvironment{syntax}[1][]{%
\(
  \begin{array}[t]{#1l@{\quad\!\!}*3{l@{}}@{\,}l}
}{
\end{array}
\)%
}

\newcommand\gdefinedby{::=}
\newcommand\gor{\mathrel{\lvert}}

\newcommand\vor{\mathrel{\big\lvert}}

\newcommand{\dif}{\mathop{}\!\mathrm{d}}

\newcommand{\Diff}{\mathbf{Diff}}
\newcommand{\wDiff}{\mathbf{\omega Diff}}

\newcommand{\wCpo}{\mathbf{\boldsymbol\omega CPO}}

\newcommand\IntwCpo[1]{\wCpo(#1)}
\newcommand\continuum{\mathfrak{c}}
\newcommand\suc[1]{#1^{+}}

\newcommand{\wds}{$\omega$-ds}
\newcommand\wdiff{$\omega$-diffeological space}
\newcommand{\wcpo}{$\omega$-cpo}
\newcommand{\wcpos}{$\omega$-cpos}
\newcommand{\wchain}{$\omega$-chain}
\newcommand{\wchains}{$\omega$-chains}
\newcommand{\wcont}{$\omega$-continuous}

\newcommand\prj[1]{#1^{\mathrm{p}}}
\newcommand\emb[1]{#1^{\mathrm{e}}}

\newcommand\ep[1]{#1^{\mathrm{ep}}}
\newcommand\embedding{\hookrightarrow}

\newcommand\lub{\bigvee}
\newcommand\rlub\bigsqcup
\newcommand\meet\wedge
\newcommand\Meet\bigwedge

\makeatletter
\def\MTrightharpoonupfill{%
  \arrowfill@\relbar\relbar\rightharpoonup}
\def\MTleftharpoondownfill{%
  \arrowfill@\leftharpoondown\relbar\relbar}
\def\MTleftharpoonupfill{%
  \arrowfill@\leftharpoonup\relbar\relbar}
\def\MTrightharpoondownfill{%
  \arrowfill@\relbar\relbar\rightharpoondown}
\newcommand*\xhookrightleftharpoons[2][]{\mathrel{%
  \raise.22ex\hbox{%
    $\lhook\joinrel\ext@arrow 0359\MTrightharpoonupfill{\phantom{#1}}{#2}$}%
  \setbox0=\hbox{%
    $\ext@arrow 3095\MTleftharpoondownfill{#1}{\phantom{\lhook\joinrel#2}}$}%
  \kern-\wd0 \lower.22ex\box0}}
\newcommand*\xleftrighthookharpoons[2][]{\mathrel{%
  \raise.22ex\hbox{%
    $\ext@arrow 3095\MTleftharpoonupfill{\phantom{#1\mspace{15mu}}}{#2}$}%
  \setbox0=\hbox{%
    $\mathrel{\raise-.4837ex\hbox{$\lhook$}}\joinrel\ext@arrow 0359\MTrightharpoondownfill{#1}{\phantom{#2}}$}%
  \kern-\wd0 \lower.22ex\box0}}
\makeatother
\newcommand\epto{\xepto{\hphantom{a}}}
\newcommand\peto{\xpeto{\hphantom{a}}}
\newcommand\xepto{\xhookrightleftharpoons}
\newcommand\xpeto{\xleftrighthookharpoons}
\WithSuffix\newcommand\pair-[2]{(#1, #2)}
\usepackage[bbgreekl]{mathbbol}

\newcommand{\Open}{\mathbf{Open}}

\newcommand{\Close}{\mathrm{Cl}}

\newcommand\unit\star

\newcommand{\Sh}{\mathbf{Sh}}

\newcommand{\CartSp}{\mathbf{CartSp}}
\newcommand{\Man}{\mathbf{Man}}
\newcommand{\CTop}{\mathbf{Top}}
\newcommand\pMap[1]{\mathbf{p}{#1}}

\usepackage{oksemantics}

\makeatletter
\DeclareRobustCommand{\overleftharpoon}{\mathpalette{\overarrow@\leftharpoonfill@}}
\DeclareRobustCommand{\overrightharpoon}{\mathpalette{\overarrow@\rightharpoonfill@}}
\def\leftharpoonfill@{\arrowfill@\leftharpoondown\mn@relbar\mn@relbar}
\def\rightharpoonfill@{\arrowfill@\mn@relbar\mn@relbar\rightharpoonup}

\DeclareMathSymbol{\leftharpoonup}{\mathrel}{okMnSymbolArrows}{'112}
\DeclareMathSymbol{\mn@relbar}{\mathrel}{okMnSymbolArrows}{'320}
\makeatother

\usepackage[normalem]{ulem}
\usepackage{contour}
\contourlength{0.8pt}

\newcommand\cnst[1]{\underline{#1}}
\def\pexplen{0mu}
\def\pexpmidlen{4mu}
\newcommand\pexp{
    \overrightharpoon{%
      {\mspace{-\pexplen}\smash{\mathord{\relbar\mspace{-\pexpmidlen}\relbar}}\vphantom{.}}\mspace{-\pexplen}}%
  }%

\newcommand\shortiff\Leftrightarrow

\DeclareMathDelimiter{\ulcorner}{\mathopen}{okMnSymbolLargeSymbols}{'036}{okMnSymbolLargeSymbols}{'036}
\DeclareMathDelimiter{\urcorner}{\mathclose}{okMnSymbolLargeSymbols}{'043}{okMnSymbolLargeSymbols}{'043}
\DeclareMathDelimiter{\llcorner}{\mathopen}{okMnSymbolLargeSymbols}{'050}{okMnSymbolLargeSymbols}{'050}
\DeclareMathDelimiter{\lrcorner}{\mathclose}{okMnSymbolLargeSymbols}{'055}{okMnSymbolLargeSymbols}{'055}
\DeclareMathDelimiter{\ullcorner}{\mathopen}{okMnSymbolLargeSymbols}{'062}{okMnSymbolLargeSymbols}{'062}
\DeclareMathDelimiter{\ulrcorner}{\mathclose}{okMnSymbolLargeSymbols}{'067}{okMnSymbolLargeSymbols}{'067}

\newcommand\constantly\underline

\newcommand\Opens[1]{\Ocal_{#1}}
\WithSuffix\newcommand\inv+[1]{\parent{#1}^{-1}}

\newcommand\Lift[1]{{#1}_{\bot}}
\newcommand\RLift[1]{{#1}_{\tilde{\bot}}}
\newcommand\Rreturn[1][{}]{\tilde{\rm return}_{#1}}
\newcommand\Rmu{\tilde{\mu}}

\newcommand{\sem}[1]{\llbracket #1\rrbracket}
\newcommand{\semgl}[1]{\llparenthesis #1\rrparenthesis}
\newcommand{\RR}{\mathbb{R}}
\newcommand{\NN}{\mathbb{N}}

\usepackage{amsmath}

\newcommand\equivalent{\simeq}
\newcommand\E{\Cat{E}_{q}}
\newcommand\wE{\Cat{E}_{\omega q}}

\newcommand\DIFF[1]{#1^\sharp}
\newcommand\invDIFF[1]{#1^\flat}
\newcommand\wDIFF[1]{#1^\sharp}

\newcommand\Sig\Sigma
\newcommand\Pres{\mathcal P}
\newcommand\pospres{\mathbf{pos}}
\newcommand\wcpopres{\boldsymbol{\omega}\mathbf{cpo}}
\newcommand\qbspres{\mathbf{qbs}}
\newcommand\diffpres{\mathbf{diff}}

\newcommand\wdiffpres{\boldsymbol{\omega}\mathbf{diff}}
\newcommand\Sorts{\mathcal S}
\newcommand\pSig{\Sig_{\mathrm{p}}}
\newcommand\tSig{\Sig_{\mathrm{t}}}
\newcommand\uSig{\Sig}
\newcommand\Eq{\mathrm{Eq}}
\newcommand\Ops{\mathcal O}
\newcommand\Var{\mathbb V}
\newcommand\arity{\mathop{\mathrm{arity}}\nolimits}
\newcommand\Term{\mathop{\mathrm{Term}}\nolimits}
\newcommand\elem{\mathrm{element}}
\newcommand\ineq{\mathrm{inequation}}

\newcommand\cover{\mathcal{U}}
\newcommand\plot{\mathrm{plot}}
\newcommand\newop[2]{\newcommand{#1}{\mathop{\mathrm{#2}}\nolimits}}

\newop\Def{Def}
\newop\source{lower}
\newop\target{upper}
\newop\irrelevance{irrel}
\newop\refl{refl}
\newop\antisym{antisym}
\newop\trans{trans}
\newop\ub{ub}
\newop\least{least}
\newop\DomainName{Dom}

\newop\ev{ev}
\newop\const{const}
\newop\rearrange{rearrange}
\newop\match{match}
\newop\extensionality{ext}

\makeatletter
\newcommand\cat[1][1]{\Cat{\@CatAlph#1}}
\newcommand\@CatAlph[1]{%
\ifcase #1\or C\or D\or A\or B\or E\else \@ctrerr \fi%
}

\newcommand\s[1][1]{%
\ifcase #1\or s\or t\or u\or v\or w\else \@ctrerr \fi%
}

\newcommand\Alg[1][1]{%
\ifcase #1\or A\or B\or C\or D\else \@ctrerr \fi%
}

\newcommand\ir[1][1]{\monad{\uir[#1]}}

\newcommand\uir[1][1]{\mathop{{}\@IRAlph{#1}}\nolimits}
\newcommand\@IRAlph[1]{%
\ifcase #1\or T\or S\or R\else \@ctrerr \fi%
}
\newcommand\grothendieck[1][1]{\Cat{\@GrothendieckAlph#1}}
\newcommand\@GrothendieckAlph[1]{%
\ifcase #1\or {J}\or {K}\else \@ctrerr \fi%
}
\makeatother
\newcommand\catC{\Cat{C}}
\newcommand\catD{\Cat{D}}
\newcommand\catE{\Cat{E}}
\newcommand\catM{\Cat{M}}
\newcommand\catV{\Cat{V}}

\newcommand\monad[1]{#1}

\newcommand\sIt{\mathrm{iterate}}
\newcommand\iterate[1]{\sIt(#1)}
\newcommand\sSign{\sem{\tSign}}
\newcommand\sInl{\mathrm{inl}}
\newcommand\sInr{\mathrm{inr}}

\newcommand\semroll[1][]{\mathrm{roll}_{#1}}

\newcommand{\bind}{\mathrel{\scalebox{0.8}[1]{\(>\!\!>\!=\)}}}
\newcommand{\klcomp}{\mathrel{\scalebox{0.8}[1]{\(>\!=\!>\)}}}
\newcommand{\return}[1][{}]{{\rm return}_{#1}}

\newcommand{\strength}{\mathrm{st}}
\newcommand{\doubstrength}{\mathrm{dst}}
\newcommand{\dist}{\mathrm{dist}}
\newcommand{\diag}{\mathrm{diag}}
\def\ci{\perp\!\!\!\perp}
\newcommand{\shLift}[1]{#1_{\ci}}

\newcommand\Dsemsymbol[1][]{\mathcal{T}^{#1}}
\newcommand\Dsem[2][]{\Dsemsymbol[#1](#2)}

\newcommand\evRsymbol[1][]{\mathrm{evR}^{#1}}
\newcommandtwoopt\evR[3][][]{\evRsymbol[#2]_{#1}(#3)}
\newcommand\lamRsymbol[1][]{\mathrm{lamR}^{#1}}
\newcommandtwoopt\lamR[3][][]{\lamRsymbol[#2]_{#1}(#3)}

\newcommand{\Gl}[1][]{\mathbf{Gl}_{#1}}
\newcommand{\wGl}[1][]{\mathbf{\omega{}Gl}_{#1}}

\newcommand{\Ocal}{\mathcal{O}}

\newcommand{\sPair}[2]{( #1, #2 )}

\newcommand\projf{\mathrm{proj}}

\makeatletter
\let\llangle\@undefined
\let\rrangle\@undefined
\DeclareMathDelimiter{\llangle}{\mathopen}%
                     {okMnSymbolLargeSymbols}{'164}{okMnSymbolLargeSymbols}{'164}
\DeclareMathDelimiter{\rrangle}{\mathclose}%
                     {okMnSymbolLargeSymbols}{'171}{okMnSymbolLargeSymbols}{'171}
\makeatother

\newcommand\glue{\mathrm{glue}}

\newcommand\DtoT[2][]{\phi_{#2#1}^{\Dsynsymbol\Dsemsymbol}}

\newcommand\sSnd{\mathrm{snd}}
\newcommand\y{\mathbf{y}}

\renewcommand\seq[2][]{\left(#2\right)_{#1}}
\renewcommand\coseq[2][]{\left[#2\right]_{#1}}
\newcommand\carrier[1]{\left\lvert#1\right\rvert}

\renewcommand\Domain[1]{\mathop{\rm Dom}\parent{#1}}

\renewcommand\lim{\mathrm{lim}}

\newcommand\ob[1]{\mathrm{ob}\,#1}
\renewcommand\circ{
  \PackageError{Matthijs}{Please use ; instead! Or if you'd rather use circ consistently, tell me and we'll switch the whole document over}{}}

\newcommand{\defeq}{\stackrel {\mathrm{def}}=}

\usepackage{todonotes}

\makeatletter
\RequirePackage{zref}

\zref@newprop{oktheoremfreetext}{\oktheorem@parameter}

\newcommand\OKTheoremAddReferences[2]{
  \expandafter\newcommand\csname#1ref\endcsname[1]{#2~\ref{#1:##1}}
  \expandafter\newcommand\csname#1label\endcsname[1]{\label{#1:##1}}
  \WithSuffix\expandafter\newcommand\csname#1ref\endcsname*[1]{\ref{#1:##1}}
  \WithSuffix\expandafter\newcommand\csname#1label\endcsname+[1]{\hypertarget{#1+:##1}{}\zref@labelbyprops{#1:##1}{oktheoremfreetext}}
  \WithSuffix\expandafter\newcommand\csname#1ref\endcsname+[1]{\hyperlink{#1+:##1}{{{\let\ref\@refstar#2~\zref@extract{#1:##1}{oktheoremfreetext}}}}}
  \WithSuffix\expandafter\newcommand\csname#1ref\endcsname-[1]{\hyperlink{#1+:##1}{{\let\ref\@refstar{\zref@extract{#1:##1}{oktheoremfreetext}}}}}
}
\makeatother
\theoremstyle{definition}

\newenvironment{tightitemize}{\begin{itemize}[leftmargin=*,noitemsep,topsep=0pt]}{\end{itemize}\ignorespacesafterend}

\newcommand\Mod[2]{#1\mathrm{-}\mathbf{Mod}(#2)}

\makeatletter
\newtheorem*{rep@theorem}{\rep@title}
\newcommand{\newreptheorem}[2]{%
\newenvironment{rep#1}[1]{%
 \def\rep@title{#2 ##1}%
 \begin{rep@theorem}\def\@currentlabel{##1}}%
 {\end{rep@theorem}}}
\makeatother

\newreptheorem{theorem}{Theorem}
\newreptheorem{corollary}{Corollary}
\newreptheorem{proposition}{Proposition}
\newreptheorem{lemma}{Lemma}

\newcommand\ConferenceArxiv[2]{#1}

\ConferenceArxiv{\AtBeginDocument{\newlabel{TotPages}{{29}{29}{}{page.29}{}}}}{}

\begin{document}

\title[Correctness of Forward AD for Iteration and Recursion]{Denotational Correctness of Forward-Mode Automatic Differentiation for Iteration
and Recursion}         


\author{Matthijs V\'ak\'ar}
\affiliation{
  \department{Department of Information and Computing Sciences}              
  \institution{Utrecht University}            
  \country{Netherlands}                    
}


\begin{abstract}
We present semantic correctness proofs of forward-mode Automatic
Differentiation (AD) for languages with sources of partiality 
such as partial operations, lazy conditionals on real parameters, iteration, and term and type recursion.
We first define an AD macro on a standard call-by-value language with some primitive operations 
for smooth partial functions and constructs for real 
conditionals and iteration, as a unique structure 
preserving macro determined by its action on the primitive operations.
We define a semantics for the language in terms of diffeological spaces,
where the key idea is to make use of a suitable partiality monad.
A semantic logical relations argument, constructed through a subsconing construction 
over diffeological spaces, yields a correctness proof of the defined AD macro.
A key insight is that, 
to reason about differentiation at sum types, we work with relations which form sheaves.
Next, we extend our language with term and type recursion.
To model this in our semantics, we introduce a new notion of space, suitable 
for modeling both recursion and differentiation, by equipping a diffeological space 
with a compatible \wcpo{-}structure. 
We demonstrate that our whole development extends to this setting.
By making use of a semantic, rather than syntactic, logical relations argument, we 
circumvent the usual technicalities of logical relations techniques for type recursion.
\end{abstract}

\ccsdesc[500]{Theory of computation~Denotational semantics}
\ccsdesc[500]{Theory of computation~Categorical semantics}
\ccsdesc[500]{Mathematics of computing~Differential calculus}
\ccsdesc[500]{Software and its engineering~General programming languages}

\keywords{automatic differentiation, semantic correctness, sconing}  

\maketitle

\section{Introduction}\label{sec:introduction}
Virtually every application of machine learning \cite{abadi2016tensorflow},
computational statistics\linebreak \cite{carpenter2017stan}
and scientific computing \cite{hascoet2013tapenade} requires efficient calculation of derivatives,
as derivatives are used in optimization, Markov integration and simulation algorithms.
Automatic Differentiation (AD) is typically the method of choice 
for algorithmically computing derivatives of programs operating between 
high-dimensional spaces, because of its efficiency and numerical stability.
In essence, AD calculates the derivative of a function implemented by 
a program by applying the chain rule across the program code.
This paper contributes a step towards making that informal statement precise,
extending the work of \cite{hsv-fossacs2020} to account for various sources 
of partiality, such as partial operations like $\log$, iteration, and recursion.

Roughly speaking, AD comes in two main flavours: forward-mode and reverse-mode.
In their naive implementation, reverse-mode outperforms forward-mode when 
calculating derivatives of functions $\RR^n\to\RR^m$
when $n\gg m$ and the other way around if $n\ll m$.
This makes reverse-mode the tool of choice, for example, when optimizing 
a real-valued objective function such as the loss of a neural network.
However, recently, \cite{shaikhha2019efficient} showed that simple forward-mode 
implementations can outperform state-of-the-art reverse-mode implementations even on such 
optimization tasks, when they are combined with standard compiler optimization techniques.
Moreover, forward-mode AD is easier to phrase and implement, and understanding how it operates on 
language features can be an important stepping stone for later grasping the complexities of advanced
reverse-mode algorithms.
Finally, there are applications where forward-mode AD is the algorithm of choice, like
in calculating dense Jacobians or Jacobian-vector products, e.g. for use in Newton-Krylov methods \cite{knoll2004jacobian}.
These methods are popular techniques for solving systems of non-linear algebraic equations,
a problem that is ubiquious in computational~physics.

In this paper, we focus on how forward-mode AD should operate on code involving the 
fundamental programming techniques of \emph{lazy conditionals}, \emph{iteration},
and \emph{(term and type) recursion}.
Our analysis works in the presence of further features of higher-order functions and product and 
sum types.
We phrase forward AD as a source-code transformation on a language with these features and 
give a proof that this transformation computes the derivatives in the usual mathematical 
sense.
Our work answers a call by the machine learning community for better developed and understood AD techniques for
expressive programming languages with features like higher-order functions and recursion \cite{van2018automatic,jeong2018improving}.
In response, we provide an understanding of forward AD on a large fragment of 
real-world functional languages such as Haskell and O'Caml.

Conditionals on real numbers are useful for pasting together functions implemented 
by language primitives.
In order to use such functions in, for example, gradient-based optimization algorithms, we need to know how to 
perform AD on these pasted functions.
For example, the ReLU function
$$
\mathbf{ReLU}(\var)\defeq \toin{cond}{(\var<\cnst{0})}{\ifelse{cond}{\tReturn\cnst{0}}{\tReturn \var}}
$$
is frequently used in the construction of neural networks.
The question of how to Automatically Differentiate such functions with "kinks" has long been studied
\cite{beck1994if}.
The current solution, like the one employed in \cite{abadi-plotkin2020}, is to treat 
such functions as undefined at their kink (in this case at $x=\cnst{0}$).
It is then up to the consumer of the differentiated code to ensure that this undefined 
behaviour does not cause problems.
This technique is used, for example, in probabilistic programming to paste together different approximations 
to a statistically important density function, to achieve numerical stability in different regimes \cite{betancourt_2019}.

Similarly, iteration constructs, or while-loops, are necessary for implementing iterative algorithms 
with dynamic stopping criteria. Such algorithms are frequently used in 
programs that need to be differentiated. For example, iteration can be used 
to implement differential equation solvers, which are routinely used (and AD'ed!)
in probabilistic programs for modelling pharmacokinetics
\cite{tsiros2019population}.
Other frequently used examples of iterative algorithms that need to be AD'ed 
are eigen-decompositions and algebraic equation solvers, such as those employed in Stan \cite{carpenter2017stan}.
Furthermore, iteration is a popular technique for achieving numerically stable approximations 
to statistically important density functions, such as that of the Conway-Maxwell-Poisson distribution \cite{goodrich_2017},
where one implements the function using a 
Taylor series, which is truncated once the next term in the series causes underflow.
For example, if we have a function whose $i$-th terms in the Taylor expansion can be represented by 
computations $i:\nat,\var:\reals\vdash^c \trm(i,\var):\reals$, we would define the underflow-truncated Taylor series by
$$
\tItFrom{
\begin{array}{l}
\pMatch{\var}{\var_1}{\var_2}{}\\
\toin{\var[2]}{\trm(\var_1,\var_2)}{}\\
\toin{\var[3]}{(\cnst{-c}<\var[2]<\cnst{c})}{}\\
\ifelse{\var[3]}{(\tReturn\,(\tInr\,\var_2))}{(\tReturn\,(\tInl\,\tPair{\var_1+\cnst{1}}{\var_2+\var[2]}))}
\end{array}
}{\var}{\tPair{\cnst{0}}{\cnst{0}}},
$$
where $\cnst{c}$ is a cut-off for underflow. 

Next, as a use case of AD applied to recursive programs, recursive neural networks 
\cite{tai2015improved} are often mentioned.
While basic Child-Sum Tree-LSTMs can also be implemented with primitive recursion (a fold) over 
an inductively defined tree, there are other related models such as Top-Down-Tree-LSTMs which require an iterative or general recursive approach \cite{zhang2016top}, where \cite{jeong2018improving} has shown that a recursive approach is preferable as it naturally exposes the available parallelism in the model.

Finally, we imagine that coinductive types like streams of real numbers, 
which can be encoded using recursive types as $\rec{\tvar}{\Unit\To(\reals\t* \tvar)}$, provide a useful API for on-line 
machine learning applications \cite{shalev2012online}, where data is processed in real time as it becomes available.
For all aforementioned applications, we need an understanding of how to perform AD on recursive programs.
This paper provides such an understanding for forward-mode AD.

\section{Key Ideas}\label{sec:key-ideas}
In this section, we give a brief conceptual summary of the key 
ideas presented in the paper.

We start off by considering a standard higher-order (fine-grain) call-by-value 
language with product and sum types over a ground type $\reals$ of real numbers and collections
$\seq[n\in\NN]{\Op_n}$ of $n$-ary basic operations. We think of these operations 
as partial functions $\RR^n\rightharpoonup \RR$ with an open domain of definition, 
on which they are smooth ($C^\infty$).
We then add the following features:
\begin{itemize}
\item (purely functional) iteration: given a computation $\Gamma,\var:\ty\vdash^c \trm:\ty\t+\ty[2]$ to iterate
and a starting value $\Gamma\vdash^v\val :\ty$, we have a computation $\Gamma\vdash^c\tItFrom{\trm}{\var}{\val}:\ty[2]$
which repeatedly calls $\trm$, starting from $\val$ until the result lies in $\ty[2]$;
\item real lazy conditionals: we assume the existence of a construct $\var:\reals\vdash^c\tSign(\var): \Unit\t+\Unit$, which lets us define 
$\ifelse{\val}{\trm}{\trm[2]}\defeq
\toin{\var}{\tSign(\val)}{
\vMatch{\var}{{
   \_\To{\trm}
\vor \_\To{\trm[2]}
}}}$.
\end{itemize}

Let us choose, for all $n\in\NN$, for all $1\leq i \leq n$, 
computations $\var_1:\reals,\ldots,\var_n:\reals\vdash^c \partial_i\op(\var_1,\ldots,\var_n):\reals$,
which represent the $i$-th partial derivative of $\op$.
For example, for the operation $\op=(*)$ of multiplication, we can choose $\partial_1(*)(\var_1,\var_2)\defeq \tReturn \var_2$.
We can then define, using fresh identifiers for all newly introduced variables,
type-respecting forward-mode AD macros $\DsynV{-}$ on values and $\DsynC{-}$ on computations, which agree on their action 
$\Dsyn{-}$ on types:\vspace{-4pt}
\begin{align*}
    &\Dsyn{\reals}\defeq\reals\times \reals\\
    &\hspace{-6pt}\begin{array}{ll}
        \DsynC{\op(\val_1,\ldots,\val_n)}\defeq~
                           &\pMatch{\DsynV{\val_1}}{\var_1}{\var_1'}
                           { \ldots \to\pMatch{\DsynV{\val_n}}{\var_n}{\var_n'}
                           {\\
                           &\toin{\var[2]}{\op(\var_1,\ldots,\var_n)}{} 
                           \\
                           &\toin{\var[3]_1}{\partial_1\op(\var_1,\ldots,\var_n)}{\ldots}\toin{\var[3]_n}{\partial_n\op(\var_1,\ldots,\var_n)}{}\\
                           &\tReturn{\tPair{\var[2]}{\var_1' *\var[3]_1+\ldots
                           +\var_n' *\var[3]_n}}}}.
    \end{array}\\
    &\DsynC{\tSign(\val)}\defeq~\pMatch{\DsynV{\val}}{\var}{\_}{\tSign(\var)}\vspace{-4pt}
\end{align*}
We then extend these definitions to a unique structure preserving macro, which acts as
$\DsynV{-}$ on values and $\DsynC{-}$ on computations.
For the cognoscenti, $(\DsynV{-},\DsynC{-})$ is structure preserving in the sense that it is a
endomorphism of 
distributive-closed Freyd categories with iteration on the syntactic category $\SynV\embed \SynC$ 
of our language (which acts on iteration as $\DsynC{\tItFrom{\trm}{\var}{\val}}=\tItFrom{\DsynC{\trm}}{\var}{\DsynV{\val}}$).
We see that the induced rule for differentiating real conditionals is 
$\DsynC{\ifelse{\val}{\trm}{\trm[2]}}=\pMatch{\DsynV{\val}}{\var}{\_}{\ifelse{\var}{\DsynC{\trm}}{\DsynC{\trm[2]}}}$.

The first-order fragment of our language has a natural semantics $\sem{-}$ in terms of 
plain multivariate calculus once we choose interpretations $\sem{\op}:\RR^n\rightharpoonup\RR$ for each $\op\in\Op_n$.
We interpret
\begin{itemize}
    \item types $\ty$ as countable disjoint unions of Euclidean spaces, $\sem{\ty}=\coprod_{i\in I}\RR^{n_i}$ (certain very simple 
    manifolds of varying dimension);
    \item values $\var_1:\ty_1,\ldots,\var_n:\ty_n\vdash^v\val :\ty[2]$ as (total) smooth functions
    $\sem{\ty_1}\times \ldots \times\sem{\ty_n}\to \sem{\ty[2]}$ between these spaces, where smoothness means that 
    the restriction to any connected component is differentiable in the usual calculus sense;
    \item computations $\var_1:\ty_1,\ldots,\var_n:\ty_n\vdash^c\trm :\ty[2]$ as partial functions
    $\sem{\ty_1}\times \ldots \times\sem{\ty_n}\rightharpoonup \sem{\ty[2]}$ between these spaces, which have an 
    open domain of definition on which they are smooth.
\end{itemize}
We can now state correctness of the macro, where we write $\Dsemsymbol_x f v$ for the usual multivariate calculus 
derivative of $f$ at $x$, evaluated on a tangent vector $v$.
\begin{theorem*}[Correctness of Fwd AD, Thm. \ref{thm:fwd-cor-full}]
For any $\var_1:\ty_1,\ldots,\var_n:\ty_n\vdash^c\trm :\ty[2]$, where $\ty_i,\ty[2]$ are first-order types,
we have that $\sem{\DsynC{\trm}}(x, v)=(\sem{\trm}(x), \Dsemsymbol_{x}\sem{\trm}v)$, for all $x$ in the domain of 
$\sem{\trm}$ and  tangent vectors $v$ at $x$.
Moreover, $\sem{\DsynC{\trm}}(x,v)$ is defined iff $\sem{\trm}(x)$ is.
\end{theorem*}
The proof follows by a straightforward induction on the structure of values and computations 
when we only consider the first-order fragment of our language.
However, we establish this theorem as well for programs between first-order types \emph{which
may include higher-order subprograms}.

To do so, we must extend our semantics $\sem{-}$ to account for higher-order programs,
which we achieve by using diffeological spaces \cite{iglesias2013diffeology}.
Diffeological spaces form a conservative extension of the usual 
setting of multivariate calculus and manifold geometry that models richer types, such as 
higher-order types.
There are many other such convenient settings for differential geometry such as Fr\"olicher
spaces \cite{frolicher1982smooth} and synthetic differential geometry \cite{kock2006synthetic},
but diffeological spaces 
to us seems like the simplest suitable setting for our purposes.

The structure of a diffeological space $X$ is a set $|X|$ together with a set $\plots{X}^U$ of functions 
$U\to X$ for any open subset $U$ of $\RR^n$ for some $n$, called the plots, which we think of 
as ``smooth functions into the space''.
Thus, we determine the geometry of the spaces by choosing the plots.
These plots need to satisfy three axioms: (1) any constant function is a plot, (2) precomposition with any smooth function 
$f:U'\to U$ in the usual calculus sense sends plots to plots, (3) we can glue compatible families 
of plots along open covers.
A homomorphism of diffeological spaces $X\to Y$, also called a smooth function, is a function 
$f:|X|\to|Y|$ such that postcomposition with $f$ maps plots to plots.
Any open subset of a Euclidean space, and more generally, any manifold, defines a diffeological space 
by taking the usual smooth functions as plots.
The smooth functions between such spaces considered as manifolds coincide with the diffeological space homomorphisms.

Next, we demonstrate how smooth partial functions with a open domain of definition are captured 
 by an easy-to-define partiality monad $\Lift{(-)}$ on the category $\Diff$ of diffeological spaces,
which lifts the usual partiality monad on the category of sets and functions: define $|\Lift{X}|\defeq |X|+\set{\bot}$
and
$$
\plots{\Lift{X}}^U\defeq \set{\alpha:U\to |X|+\set{\bot}\mid \alpha^{-1}(X)\subseteq_{\mathrm{open}}U\textnormal{ and } \alpha|_{\alpha^{-1}(X)}\in \plots{X}^{\alpha^{-1}(X)} }.
$$
Indeed, total functions $M\to \Lift{N}$ in case $M$ and $N$ are manifolds correspond precisely 
to partial functions $M\rightharpoonup N$ that have an open domain of definition in the Euclidean topology 
and that are smooth on that domain.
We show that on general diffeological spaces, this monad classifies smooth partial functions that have a domain 
of definition that is open in the well-studied D-topology \cite{christensen2014d}.
Moreover, we show that this is a commutative strong monad that models iteration: it is 
a complete Elgot monad in the sense of \cite{goncharov2015unguarded}.
The idea is to interpret $\tItFrom{\trm}{\var}{\var[2]}$ as the union (lub) over $i$ of the $i$-fold 
``self-compositions'' of $\sem{\trm}$. 
This monad is also easily seen to interpret $\tSign$ as the smooth partial function $\sem{\tSign}:\RR\to\terminal + \terminal$ 
that sends the positive reals to the left copy of $\terminal$ and the negative reals to the right.
As diffeological spaces form a bicartesian closed category, we now obtain a canonical interpretation 
$\sem{-}$ of our entire language. We interpret
\begin{itemize}
\item types $\ty$ as diffeological spaces;
\item values $\var_1:\ty_1,\ldots,\var_n:\ty_n\vdash^v \ty[2]$ as smooth functions $\sem{\ty_1}\times\ldots\times\sem{\ty_n}\To\sem{\ty[2]}$;
\item computations $\var_1:\ty_1,\ldots,\var_n:\ty_n\vdash^c\trm :\ty[2]$ as smooth functions
$\sem{\ty_1}\times \ldots \times\sem{\ty_n}\to \Lift{\sem{\ty[2]}}$, or, equivalently,
as partial functions $\sem{\ty_1}\times \ldots \times\sem{\ty_n}\rightharpoonup {\sem{\ty[2]}}$
that, on their domain, restrict to smooth functions $U\to\sem{\ty[2]}$ where $U$ is an open subset in the D-topology of $\sem{\ty_1}\times \ldots \times\sem{\ty_n}$.
\end{itemize}

To establish the correctness theorem for the full language, we use a logical relations argument 
over the semantics in diffeological spaces, where we maintain a binary relation that relates smooth curves 
$\RR\to\sem{\ty}$ to their tangent curve $\RR\to\sem{\Dsyn{\ty}}$.
We derive this argument by using a subsconing construction diffeological spaces, like the one that is 
employed in \cite{hsv-fossacs2020}.
The key step is to find a suitable definition of the partiality monad $\Lift{(-)}$ on relations as 
discussed in \cite{goubault2002logical}.
This lifting can be achieved by working with relations that are local in the sense that global membership 
of the relation can be restricted to membership on subsets of the domain. Conversely, 
if we establish local membership of the relation, we can derive global membership by gluing.
Put differently, we work with \emph{relations that are sheaves} over open subsets of $\RR$.
This approach is justified because differentiation is local operation, and we can glue smooth 
functions and their derivatives.
Once we work with such a relation $\seq[U\subseteq_{\mathrm{open}} \RR]{R^U\subseteq (U\To X)\times (U\To Y)}$, we can simply define
the relation $\RLift{R}^U\subseteq (U\To \Lift{X})\times (U\To\Lift{Y})$ as containing 
those pairs $(\gamma,\gamma')$ of partially defined curves that have the same domain of definition $V\subseteq_{\mathrm{open}}U$
and which are in the relation on this domain: $(\gamma|_{V},\gamma'|_{V})\in R^V$.
The correctness theorem then follows by standard logical relations (subsconing) techniques.
We note that working with relations that are sheaves is particularly important to establish correctness 
for sum types and $\ifelse{-}{-}{-}$.

While category theory helped us find this proof, it can be phrased entirely in elementary terms.
Indeed, we define, for each open $U\subseteq \RR$, relations $R^U_{\ty}\subseteq (U\To\sem{\ty})\times (U\To\sem{\Dsyn\ty})$
and $\RLift{(R^U_{\ty})}\subseteq (U\To\Lift{\sem{\ty}})\times (U\To\Lift{\sem{\Dsyn{\ty}}})$, using induction on the types $\ty$ of our language:
\begin{align*}
R^U_{\reals}&\defeq \set{\sPair{\gamma}{\gamma'}\mid \forall x\in U. \gamma'(x)=\left(\gamma(x), \frac{\dif}{\dif t}|_{t=x}\gamma(t)\right)}\\
R^U_{\ty_1\t*\ty_2} &\defeq \set{\sPair{\gamma}{\gamma'}\mid \sPair{\gamma;\pi_1}{\gamma';\pi_1}\in R^U_{\ty_1}\textnormal{ and }
\sPair{\gamma;\pi_2}{\gamma';\pi_2}\in R^U_{\ty_2}}\\
R^U_{\ty_1 \t+ \ty_2} &\defeq \set{\sPair{\gamma}{\gamma'}\mid \forall i=1,2.(\gamma|_{\gamma^{-1}(\sem{\ty_i})},\gamma'|_{\gamma^{-1}(\sem{\ty_i})})\in R^{{\gamma^{-1}(\sem{\ty_i})}}_{\ty_i}}\\
R^U_{\ty_1\To \ty_2} &\defeq \set{\sPair{\gamma}{\gamma'}\mid \forall (\delta,\delta')\in R^U_{\ty_1}. (x\mapsto \gamma(x)(\delta(x)), x\mapsto \gamma'(x)(\delta'(x)))\in \RLift{(R^U_{\ty_2})}}\\
\RLift{(R^U_{\ty})}&\defeq \set{\sPair{\gamma}{\gamma'}\mid \gamma^{-1}(\sem{\ty})=\gamma'{}^{-1}(\sem{\Dsyn{\ty}})\textnormal{ and }
\sPair{\gamma|_{ \gamma^{-1}\sem{\ty}}}{\gamma'|_{ \gamma^{-1}\sem{\ty}}}\in R^{{ \gamma^{-1}\sem{\ty}}}} 
\end{align*}
We then establish the following ``fundamental lemma'':
\begin{quote}
If $\var_1:\ty_1,\ldots,\var_n:\ty_n\vdash^v \val:\ty[2]$ (resp. $\var_1:\ty_1,\ldots,\var_n:\ty_n\vdash^c \trm:\ty[2]$)
and, 
for $1\leq i\leq n$, $(f_i,g_i)\in R^{U}_{\ty_i}$, for $U\subseteq_{\mathrm{open}}\RR$,
then we have that $\left((f_1,\ldots,f_n);\sem{\val}, (g_1,\ldots,g_n);\sem{\DsynV{\val}}\right)\in R^{U}_{\ty[2]}$
(resp.  $\left((f_1,\ldots,f_n);\sem{\trm}, (g_1,\ldots,g_n);\sem{\DsynC{\trm}}\right)\in \RLift{(R^{U}_{\ty[2]})}$).
\end{quote}
The proof follows by induction on the typing derivation of $\val$ and $\trm$, where the 
only non-trivial step is to show that each basic operation\, $\op$\, respects the relations
(and the other steps follow by standard results about logical relations/subsconing).
This basic step follows from the chain rule for differentiation, provided that the 
derivatives of the basic operations are correctly implemented:
$$
\sem{\partial_i\op(\var_1,\ldots,\var_n)}=\nabla_i \sem{\op(\var_1,\ldots,\var_n)},
$$
where we write $\nabla_i f$ for the usual calculus partial derivative of a
function $f$
in the direction of the $i$-th standard basis vector of $\RR^n$. 
The correctness result is a straightforward corollary of this~lemma.

Next, we extend our language with term and type recursion, in the sense of
FPC \cite{fiore1994axiomatisation}.
To apply the AD macro to recursive types, we first need to define its action on type variables $\tvar$:
$\Dsyn{\tvar}\defeq\tvar$.
Next, we can define its action on recursive types, the corresponding values and computations,
as well as the induced action on term recursion $\rec{\var}\trm$, which can be seen as sugar:
\begin{align*}
&\Dsyn{\rec{\tvar}{\ty}}\defeq \rec{\tvar}{\Dsyn{\ty}}\qquad\quad
\DsynV{\tRoll{\val}}\defeq \tRoll{\DsynV{\val}}\qquad\quad\DsynC{\rec{\var}\trm}\defeq \rec{\var}{\DsynC{\trm}}\\
&\DsynC{\rMatch{\val}{\var}{\trm}}\defeq \rMatch{\DsynV{\val}}{\var}{\DsynC{\trm}}.
\end{align*}
Our semantics in diffeological spaces appears not to suffice to model recursion, for the same 
reason that sets and partial functions do not.
Therefore, we follow the steps taken in \cite{vakar2019domain} and
introduce a new notion of space, suitable for modelling both recursion and differentiation:
we equip the diffeological spaces 
with a compatible \wcpo{} (chain-complete partial order) structure. 
\begin{definition*}
An $\omega$-diffeological space comprises a set $|X|$ with the following structure:
\begin{itemize}
\item a diffeology, i.e. specified sets of functions $\plots{X}^U\subseteq  U\To |X|$ for $U\subseteq_{\mathrm{open}}\RR^n$
subject to the three conditions discussed above: to enforce differentiability;
\item a partial order $\leq$ on $|X|$ such that least upper bounds (lubs) of $\omega$-chains exist: to model recursion,
\end{itemize}
such that each $\plots{X}^U$ is closed under pointwise lubs of pointwise $\omega$-chains.
\end{definition*}
Equivalently, we can define them as \wcpos{} internal to $\Diff$.
We write $\wDiff$ for the category of $\omega$-diffeological spaces and
diffeological space homomorphisms that are, additionally, $\omega$-continuous in the sense of preserving 
lubs of \wchains{}.
$\wDiff$ is again a bicartesian closed category.

We can observe that our previous definition of the partiality monad on $\Diff$ lifts to give a
partiality monad on $\wDiff$: interpret $\bot$ as a new least element in the order.
Again, we obtain a semantics $\sem{-}$ of our original language in $\wDiff$, where we equip $\RR$ with the 
discrete order structure, in which all elements are only comparable to themselves.
Moreover, we can show that the Kleisli adjunction $\wDiff\leftrightarrows \Lift{\wDiff}$
has sufficient structure to interpret term and type recursion. Indeed, it gives rise to a so-called bilimit
compact expansion \cite{levy2012call},
which follows from a development of axiomatic domain theory, similar to that in \cite{vakar2019domain}.
As a consequence, we can extend the semantics of our original language to one that also accounts for 
term and type recursion.

Finally, we extend our logical relations proof to apply to recursive structures.
Proving the existence of logical relations defined using type recursion is notoriously difficult
and much advanced machinery was developed to accomplish this feat \cite{pitts1996relational}.
Much of the difficulty is caused by the fact that one customarily considers logical relations over 
the syntax, which is not chain-complete.
Here, however, we consider relations over a chain-complete semantics.
As a consequence, we can circumvent many of the usual technicalities.
In particular, our previous logical relations proof extends if we work, instead, with relations 
that are internal \wcpos{} in the category of sheaves on open subsets of $\RR$.
We show that this category of logical relations lifts the bilimit compact expansion structure of 
$\wDiff$, meaning that the interpretation of recursive types lifts.
Stated in elementary terms, we can define logical relations
$\seq[U\subseteq_{\mathrm{open}}\RR]{R^U\subseteq (U\To \sem{\ty})\times (U\To\sem{\Dsyn{\ty}})}$
using type recursion, as along as
we work only with relations 
for which each $R^U$ is closed under lubs of $\omega$-chains 
of its elements $(\gamma,\gamma')$.
In the end, once we have suitably extended the definition of our logical relations, we establish the same fundamental lemma and
the same correctness theorem follows, but now for our more expressive language,
in which ``first-order types'' include algebraic data types like lists and trees.
\section{An AD translation for Iteration and Real Conditionals}\label{sec:language1}
\subsection{A simple call-by-value language}
We consider a standard fine-grain call-by-value language over a ground type $\reals$ of 
real numbers, real constants $\cnst{c}\in\Op_0$ for $c\in\RR$,
and certain basic operations $\op\in\Op_n$ for each natural number $n\in\NN$.
We will later interpret these operations as partial functions $\RR^n\rightharpoonup \RR$ with domains
of definition that are (topologically) open in $\RR^n$ on which they are smooth ($C^\infty$) functions.
These operations include, for example,
unary operations on reals like $\exp,\sigma,\log,\sigmoid\in \Op_1$ (where we mean the mathematical sigmoid function
$\sigmoid(x)\defeq\frac 1 {1+e^{-x}}$),
binary operations on reals like $(+),(-),(*),(/)\in\Op_2$.
We treat this operations in a schematic way as this reflects the reality of 
practical AD libraries which are constantly being expanded with new primitive operations.
We consider a fine-grain CBV language mostly because it makes the proofs more conceptually clean,
but we could have equally worked with a standard coarse-grain CBV language which can be faithfully 
embedded in our language.
We discuss the implied differentiation rules for such a language in Appx. \ref{sec:coarse-grain-cbv},
which inherit the correctness results that we will prove for the fine-grain language.

The types $\ty,\ty[2],\ty[3]$, values $\val,\val[2],\val[3]$, and computations $\trm,\trm[2],\trm[3]$ of our language are as follows.\\
\begin{syntax}
    \ty, \ty[2], \ty[3] & \gdefinedby & & \syncat{types}                          \\
    &\gor& \reals                      & \synname{numbers}\\
    &\gor & \Init \ \gor \ty \t+ \ty[2]  & \synname{sums}\\
    &&&\\
    \val, \val[2], \val[3] & \gdefinedby & & \syncat{values}                          \\
    &\gor& \var,\var[2],\var[3]                      & \synname{variables}\\
    &\gor& \cnst{c}                   & \synname{constant}\\
    &&&\\
    \trm, \trm[2], \trm[3] & \gdefinedby & & \syncat{computations}                          \\
    &\gor & \toin{\var}{\trm}{\trm[2]} & \synname{sequencing}\\
    &\gor & \tReturn \val & \synname{pure comp.}\\
    &\gor & \op(\val_1,\ldots,\val_n) & \synname{operation}\\
    &\gor & \nvMatch{\val} & \synname{sum match}\\
    &\gor & \vMatch{\val}{\begin{array}{l}\;\;\tInl\var\To\trm\\
    \gor \tInr\var[2]\To \trm[2]\end{array}}\hspace{-15pt}\; & \synname{sum match}\\
  \end{syntax}%
  ~
  \begin{syntax}
    &\gor\quad\, & \Unit \ \gor  \ty_1\t* \ty_2 & \synname{products}\\
  &\gor& \ty \To \ty[2]              & \synname{function}      \\
  & & &\\
  &&&\\
  &\gor& \tInl{\val}\ \gor   \tInr{\val} & \synname{sum inclusions}\\
  &\gor\quad\, & \tUnit \ \gor  \tPair{\val}{\val[2]} & \synname{tuples}\\
&\gor& \fun{\var}{\trm}             & \synname{abstractions}      \\
&&&\\
&\gor\quad\, & \uMatch{\val}{\trm} & \synname{product match}\\
&\gor\quad\, & \pMatch{\val}{\var}{\var[2]}{\trm} & \synname{product match}\\
&\gor& \val\ \val[2]             & \synname{function app.}      \\
&\gor&\tItFrom{\trm}{\var}{\val} & \synname{iteration}\\
&\gor&\tSign\val & \synname{sign function}
  \end{syntax}
\\
We will use sugar $\letin{\var}{\val}{\val[2]}\defeq \subst{\val[2]}{\sfor{\var}{\val}}$, 
$\letin{\var}{\val}{\trm}\defeq (\fun{\var}{\trm})\ \val$, and 
$\ifelse{\val}{\trm}{\trm[2]}\defeq \toin{\var}{\tSign(\val)}
{\vMatch{\var}{{
   \_\To{\trm[2]}
\vor \_\To{\trm[3]}
}}}$.
The typing rules are in Figure~\ref{fig:types1}.

\begin{figure}[b]
  \framebox{\scalebox{1.0}{\begin{minipage}{\linewidth}\noindent\input{type-system1}\end{minipage}}}
  \caption{Typing rules for the our fine-grain CBV language with iteration and real conditionals.
  We use a typing judgement $\vdash^v$ for values and $\vdash^c$ for computations.\label{fig:types1}}
  \end{figure}

\subsection{A translation for forward-mode Automatic Differentiation}
Let us fix, for all $n\in\NN$, for all $\op\in\Op_n$, for all $1\leq i \leq n$, 
computations $\var_1:\reals,\ldots,\var_n:\reals\vdash^c \partial_i\op(\var_1,\ldots,\var_n):\reals$,
which represent the partial derivatives of $\op$.
For example, we can choose $\partial_1(+)(\var,\var[2])\defeq \partial_2(+)(\var,\var[2])\defeq \tReturn\cnst{1}$,
$\partial_1(*)(\var,\var[2])\defeq \tReturn \var[2]$ and $\partial_2(*)(\var,\var[2])\defeq \tReturn\var[1]$,
$\partial_1\log(\var)\defeq \cnst 1/\var$ and
$\partial_1\sigmoid(\var)\defeq \toin{\var[2]}{\sigmoid(\var)}{\toin{\var[3]}{\cnst{1}-\var[2]}{\var[2]*\var[3]}}$.
Using these terms for representing partial derivatives,
we define, in Fig. \ref{fig:ad1}, a structure preserving macro $\Dsynsymbol$ on the types, values, and computations of our language for performing 
forward-mode AD.
We observe that this induces the following AD rule for our sugar: $\DsynC{\ifelse{\val}{\trm}{\trm[2]}}=\pMatch{\DsynV{\val}}{\var}{\_}{\ifelse{\var}{\DsynC{\trm}}{\DsynC{\trm[2]}}}$.
\begin{figure}[b]
   \framebox{\scalebox{1.0}{\begin{minipage}{\linewidth}\noindent
\input{d-types1}
\hrulefill
\input{d-values1}
\hrulefill
\input{d-computations1}
\end{minipage}}}
\caption{A forward-mode AD macro defined on types as $\Dsyn{-}$, values as $\DsynV{-}$, and computations as $\DsynC{-}$.
All newly introduced variables are chosen to be fresh.
We single out the rule for primitive operations $\op$, as this is where the interesting work specific to differentiation happens.\label{fig:ad1}}
\end{figure}

\subsection{Equational theory}
We consider our language up to the usual $\beta\eta$-equational theory for fine-grain CBV,
which is displayed in Fig. \ref{fig:beta-eta1}.
We could impose further equations for the iteration construct as is done in e.g. \cite{bloom1993iteration,goncharov2015unguarded}
as well as for the basic operations\, $\op$\, and the sign function $\tSign$.
However, such equations are unnecessary for our development.

\begin{figure}[b]
   \framebox{\scalebox{1.0}{\begin{minipage}{\linewidth}\noindent
\input{beta-eta1}
\end{minipage}}}
\caption{Standard $\beta\eta$-laws for fine-grain CBV.
We write $\freeeq{\var_1,\ldots,\var_n}$ to indicate that the variables are fresh in the left hand side.
In the top right rule, $\var$ may not be free in $\trm[3]$.
Equations hold on pairs of terms of the same type.\label{fig:beta-eta1}}
\end{figure}

We extend $\Dsynsymbol$ to contexts: $\Dsyn{\{\var_1{:}\ty_1,{.}{.}{.},\var_n{:}\ty_n\}}\defeq
\{\var_1{:}\Dsyn{\ty_1},{.}{.}{.},\var_n{:}\Dsyn{\ty_n}\}$.
This turns $\Dsynsymbol$ into a well-typed, functorial macro in the following sense.
\begin{lemma}[Functorial macro]\label{lem:functorial-macro}
Our macro respects typing, substitution, and $\beta\eta$-equality:
\begin{itemize}
\item If $\Ginfv\val\ty$, then $\Dsyn{\Gamma}\vdash^v\DsynV{\val}:\Dsyn{\ty}$.\quad $\bullet$ If $\Ginfc\trm\ty$, then $\Dsyn{\Gamma}\vdash^c\DsynC{\trm}:\Dsyn{\ty}$.
\item $\DsynV{\letin{\var}{\hspace{-3pt}\val}{\val[2]}}=\letin{\var}{\hspace{-3pt}\DsynV\val}{\DsynV{\val[2]}}\;\;
\DsynC{\letin{\var}{\val}{\trm}}=\letin{\var}{\hspace{-3pt}\DsynV\val}{\DsynC{\trm}}$.
\item If $\val\beeq\val[2]$, then $\DsynV{\val}\beeq \DsynV{\val[2]}$.\quad\quad\qquad $\bullet$ If $\trm\beeq\trm[2]$, then $\DsynC{\trm}\beeq\DsynC{\trm[2]}$.
\end{itemize}
\end{lemma}
   This lemma can be shown manually, or the categorically-inclined reader can observe that it follows from the 
   following universal property of the syntax.

   Indeed, we can define categories $\SynV$ and $\SynC$ which both have types as objects. 
   $\SynV(\ty,\ty[2])$ consists of $(\alpha)\beta\eta$-equivalence classes of 
   values $\var:\ty\vdash^v \val:\ty[2]$, where identities are $\var:\ty\vdash^v \var:\ty[2]$ 
   and composition of $\var:\ty\vdash^v \val:\ty[2]$ and $\var[2]:\ty[2]\vdash^v \val[2]:\ty[3]$
   is given by $\var:\ty\vdash^v \letin{\var[2]}{\val}{\val[2]}:\ty[3]$.
   Similarly, $\SynC(\ty,\ty[2])$ consists of $(\alpha)\beta\eta$-equivalence classes of 
   computations $\var:\ty\vdash^c \trm:\ty[2]$, where identities are $\var:\ty\vdash^c \tReturn\var:\ty[2]$ 
   and composition of $\var:\ty\vdash^c \trm:\ty[2]$ and $\var[2]:\ty[2]\vdash^c \trm[2]:\ty[3]$
   is given by $\var:\ty\vdash^c \toin{\var[2]}{\trm}{\trm[2]}:\ty[3]$.
   $\SynV\embed\SynC;\val\mapsto \tReturn\val$ is
   the free distributive-closed Freyd category with iteration
   generated by $\reals$, the constants $\cnst c$ and operations $\op$ and $\tSign$.
   It has the universal property that for any other distributive-closed Freyd category \cite{levy2003modelling}
   with iteration $\catV\embed\catC$, we obtain a unique structure preserving functor 
   $(F_V,F_C):(\SynV\embed \SynC)\to (\catV\embed \catC)$ once we fix an object $F(\reals)$
   and compatibly typed morphisms $F_V(\cnst c)\in \catV(\terminal, F(\reals))$,
   $F_C(\op)\in \catC(F(\reals)^n,F(\reals))$ and $F_C(\tSign)\in\catC(F(\reals), \terminal+\terminal)$.
\section{A denotational semantics via diffeological spaces}\label{sec:semantics1}

\subsection{Preliminaries}
\subsubsection*{Category theory}
We assume familiarity with categories $\cat$, $\cat[2]$, functors
$F, G : \cat \to \cat[2]$, natural transformations
$\alpha, \beta : F \to G$, and their theory of (co)limits
and adjunctions.
We write:
\begin{itemize}
\item unary, binary, and $I$-ary products
as $\terminal$, $X_1\times X_2$, and $\prod_{i\in I}X_i$, writing
$\projection_i$ for the projections and
$()$, $\sPair{x_1}{x_2}$, and $\seq[i\in I]{x_i}$ for the tupling maps;
\item unary, binary, and $I$-ary coproducts
as $\initial$, $X_1 + X_2$, and $\sum_{i\in I}X_i$, writing
$\injection_i$ for the injections and
$[]$, $[{x_1},{x_2}]$, and $\coseq[i\in I]{x_i}$ for the cotupling
maps;
\item exponentials as $X^Y$, writing $\Lambda$ for the currying
 maps.
\end{itemize}

\subsubsection*{Monads}
A \emph{strong monad} $\monad T$ over a
cartesian closed category $\cat$ is a triple $\triple T\return\bind$
consisting of an assignment of an object $TX$ and a morphism
$\return[X] : X \to TX$ for every object $X$, and an assignment of a
morphism $(\bind_{X,Y}) : TX \times (TY)^X \to TY$, satisfying the monad
laws below, expressed in the cartesian closed internal language:
\[
  \parent{(\return x) \bind f} = f(x) \qquad \parent{a \bind \return} = a
  \qquad \parent{(a \bind f) \bind g} = \parent{a \bind \lambda x. f(x) \bind g}
\]
Every monad yields an endofunctor $T$ on $\cat$-morphisms:
$T(f):=\id\bind^T(f;\return^T)$.  The \emph{Kleisli} category
$\cat_{\ir}$ consists of the same objects as $\cat$, but morphisms are
given by $\cat_{\ir}(X, Y) \defeq \cat(X, \uir Y)$. Identities are $\return$
and composition $f\klcomp g$ is given by $\lambda x. f x \bind g$. A strong monad
$\ir$ is \emph{commutative} when, for every pair of objects $X,Y$:
\[
a : \uir X,b : \uir Y\vdash
  a \bind \lambda x. b \bind \lambda y. \return (x,y)
  =
  b \bind \lambda y. a \bind \lambda x. \return (x,y)
\]
We write $\doubstrength_{X,Y}$ for this morphism and call it the \emph{double strength}.

\subsubsection*{Semantics of fine-grain CBV}
Generally, fine-grain call-by-value languages like ours can be interpreted in distributive-closed Freyd categories
(with additional support for interpreting iteration, in our case).
However, we do not need this level of generality.
Instead, we focus on the more specific semantic settings
of bicartesian closed categories equipped with suitable 
partiality monads.

Given a bicartesian closed category $\catC$ and a strong monad 
$(T,\return,\bind)$, we have a sound interpretation $\sem{-}$ of fine-grain CBV
with its $\beta\eta$-equational theory (see \cite{levy2003modelling}, for details):
\begin{align*}
&\sem{\Init} \defeq \initial\quad
\sem{\ty\t+\ty[2]} \defeq \sem{\ty}+\sem{\ty[2]}&&\hspace{-42pt}
\sem{\Unit}\defeq\terminal\quad
\sem{\ty\t*\ty[2]} \defeq \sem{\ty}\times \sem{\ty[2]}\quad
\sem{\ty\To\ty[2]} \defeq T\sem{\ty[2]}^{\sem{\ty[1]}}\\
&\sem{\var_1:\ty_1,\ldots,\var_n:\ty_n} \defeq \sem{\ty_1}\times\cdots\times\sem{\ty_n}&&\hspace{-42pt} \sem{\Gamma\vdash^v \ty}\defeq\catC(\sem{\Gamma}, \sem{\ty})\quad \sem{\Gamma\vdash^c \ty}\defeq\catC_T(\sem{\Gamma}, \sem{\ty})\\
    &
    \sem{\var_1:\ty_1,\ldots,\var_n:\ty_n\vdash^v\var_i}(\rho)\defeq \projection_i(\rho)
    &&
    \sem{\toin{\var}{\trm}{\trm[2]}}(\rho)\defeq \sem{\trm}(\rho)\bind \sem{\trm[2]}(\rho,\id) \\
    &
    \sem{\tReturn{\val}}(\rho)\defeq \return(\sem{\val}(\rho))
    &&
    \sem{\nvMatch{\val}}(\rho) \defeq [](\sem{\val}(\rho))\\
    &
    \sem{\tInl{\val}}(\rho)\defeq \injection_1(\sem{\val}(\rho))
    &&
    \sem{\tInr{\val}}(\rho)\defeq \injection_2(\sem{\val}(\rho))\\
    &
    \begin{array}{l}\hspace{-5pt}\sem{\vMatch{\val}{\tInl\var\To\trm
    \gor \tInr\var[2]\To {\trm[2]}}}(\rho)\defeq\\
\quad [\sem{\trm}(\rho,\id),\sem{\trm}(\rho,\id)](\sem{\val}(\rho))\end{array}
    &&
    \sem{\tUnit}(\rho)\defeq ()\\
    &
    \sem{\tPair{\val}{\val[2]}}\defeq \sPair{\sem{\val}(\rho)}{\sem{\val[2]}(\rho)}
    &&
    \sem{\pMatch{\val}{\var}{\var[2]}{\trm}}(\rho)\defeq \sem{\trm}(\rho,\sem{\val}(\rho))\\
    &
    \sem{\fun{\var{:}\ty}{\trm}}(\rho)(a)\defeq
    \sem {\trm}(\rho,a)\quad \text{($a\in \sem {\ty}$)}
    &&
    \sem{\trm\,\val}(\rho)\defeq
    \sem{\trm}(\rho)(\sem{\val}(\rho))\vspace{-6pt}
    \end{align*}

To interpret the constructs for iteration, we need
a chosen family of functions
$\sIt_{C,A,B}:\catC_T(C\times A,A+B)\to\catC_T(C\times A, B)$ such that the following naturality
condition holds with respect to morphisms $f:C\to C'$ in
$\catC$: $(f\times \id[A]);\sIt_{C',A,B}(g)=\sIt_{C,A,B}((f\times \id[A]);g)$.
Then, we define, for $\Ginfc[,\var:{\ty[2]}]\trm{\ty[2]\t+\ty}$
and $\Ginfv\val\ty[2]$,
$\sem{\tItFrom{\trm}{\var}{\val}}(\rho)\defeq\sIt_{\sem{\Gamma},\sem{\ty[2]},\sem{\ty}}(\sem{\trm})(\rho,\sem{\val}(\rho))$.
We say that a functor $F$ between two categories which support iteration 
preserves iteration if $\sIt(F(f))=F(\sIt(f))$.

Given bicartesian closed category $\catC$ with a strong monad $T$ 
such that $\catC_T$ supports iteration, therefore, 
we obtain a canonical interpretation $\sem{-}$ of our full language, once we 
\emph{choose} an interpretation of $\reals$, $\tSign$, constants $\cnst{c}$ and operations $\op$.
That is we need a chosen object 
$\sem{\reals}$ with chosen morphisms $\sSign\in\catC_T(\sem{\reals}, \terminal +\terminal)$,
$\sem{{c}}\in\catC(\terminal,\sem{\reals})$ and chosen morphisms $\sem{\op}\in \catC_T(\sem{\reals}^n,\sem{\reals})$ for each $\op\in\Op_n$.
We then interpret \\
$
\sem{\cnst{c}}(\rho)\defeq \sem{c}\quad\!\sem{\op(\val_1,\ldots,\val_n)}(\rho)\defeq \sem{\op}(\sem{\val_1}(\rho),\ldots,\sem{\val_n}(\rho))
\quad\!
\sem{\tSign(\val)}(\rho)\defeq \sSign(\sem{\val}(\rho)).
$
A categorically inclined reader might want to note that $\sem{-}$
is canonical as the uniquely determined structure preserving 
functor $(\SynV \embed\SynC)\to(\catC\embed\catC_T)$ of distributive-closed
Freyd categories with iteration,
which we discussed in Sec. \ref{sec:language1}.

\subsubsection*{\wcpos{}}
We recall some basic domain theory. Let $\omega=\set{0 \leq
1 \leq \ldots}$ be the ordinary linear order on the naturals. An \emph\wchain{}
in a poset $P = \pair{\carrier P}{\leq}$ is a monotone function $a_-
: \omega \to P$. A poset $P$ is an \emph{\wcpo}
when every \wchain{} $\seq[n \in \NN]{a_n}$ has a least upper bound
(lub) $\lub_{n \in \NN}a_n$ in $P$.

\begin{example}
  Each set $X$ equipped with the discrete partial order forms an \wcpo{}
  $\pair X=$, e.g., the discrete \wcpo{} $\RR$ over the
  real line.
  For any \wcpo{} $X$, we can form a new \wcpo{} $\Lift{X}$ by adding 
  a new element $\bot$ which will be the least element in the order.
  $\Lift{(-)}$ lifts the partiality monad on $\Set$.
\end{example}

For \wcpo s $P$ and $Q$, an \emph{\wcont}
function $f : P \to Q$ is a monotone function $f : \carrier P \to
\carrier Q$ such that for every \wchain{} $a_-$, we have: $f(\lub_n
a_n) = \lub_n f(a_n)$.  Such a function $f : P \to Q$ is a
\emph{full mono} when, for every $a,b \in \carrier P$, we have $f(a)
\leq f(b) \implies a \leq b$. Recall that the category $\wCpo$ of
\wcpo s and \wcont{} functions is bicartesian closed: coproducts 
are disjoint unions on both the carrier and order, products
are taken componentwise and the exponential $Q^P$ has carrier
$\wCpo(P,Q)$ and order $f\leq_{Q^P} g$ iff $\forall p\in
|P|.f(p)\leq_Q g(p)$. A \emph{pointed \wcpo{}} is an \wcpo{} with a least
element $\bot$. A \emph{strict} function is a
\wcont{} function that preserves the least elements.
The Kleisli category $\Lift{\wCpo}$ is equivalent to the category of \wcpos{} and strict \wcont{} functions.

An \emph{$\wCpo$-(enriched) category} $\cat$ consists of a
locally-small category $\cat$ together with an assignment of an
\wcpo{} $\cat(A, B)$ to every $A, B \in \ob{\smash{\cat}}$ whose carrier is
the set $\cat(A, B)$ such that composition is \wcont. An
\emph{$\wCpo$-functor}, a.k.a.~a \emph{locally-continuous functor},
$F : \cat \to \cat[2]$ between two $\wCpo$-categories is an ordinary
functor $ F : \cat \to {\cat[2]}$ between the underlying
categories, such that every morphism map
$ F_{A, B} : \cat(A, B) \to \cat[2](FA, FB)$ is \wcont.
We call such a category $\Lift{\wCpo}$-enriched if all homsets 
has a least element and composition is strict in both~arguments.

\begin{example}
  Every locally-small category is an $\wCpo$-category whose
  hom-\wcpos{} are discrete. The category $\wCpo$ itself is an
  $\wCpo$-category. If $\cat$ is an $\wCpo$-category, its categorical
  dual $\opposite\cat$ is an $\wCpo$-category. The category of
  locally-continuous functors $\opposite\cat \to \wCpo$, with the
  componentwise order on natural transformations $\alpha : F \to G$,
  is an $\wCpo$-category when $\cat$ is.
\end{example}

By the Knaster-Tarski theorem, an \wcont{} function $f:X\to X$ on a pointed 
\wcpo{} $X$~has a least fixed-point,
i.e. a least $x$ such that $x=f(x)$.
This is given by the lub of the \wchain{} $\seq[n\in\NN]{f^n(\bot)}$.

\subsubsection*{Manifolds}
For our purposes, a smooth manifold $M$ is a second-countable Hausdorff
topological space with a smooth atlas: an open cover $\cover$ together
with homeomorphisms $\seq[U\in\cover]{\phi_U:U\to \RR^{n(U)}}$ (called charts) such that $\phi_U^{-1};\phi_V$ is smooth on its domain of definition for all $U,V\in\cover$.
A function $f:M\to N$ between manifolds is smooth if $\phi^{-1}_U;f;\psi_V$ is
smooth for all charts $\phi_U$ and $\psi_V$ of $M$ and $N$, respectively.
Let us write $\Man$ for the category of smooth manifolds and smooth functions.
We are primarily interested in manifolds that are isomorphic to $\coprod_{i\in I}\RR^{n_i}$
for countable $I$, as they will form the interpretation of our first-order types.

Note that different charts in the atlas of our manifolds may have different finite dimension $n(U)$.
Thus we consider manifolds with dimensions that are potentially unbounded, albeit locally finite.
This non-constant dimension does not affect the theory of differential geometry as far as we need~it.

\subsubsection*{Diffeological spaces}
We consider a rich notion of differential geometry for our semantics, based on 
\emph{diffeological spaces} \cite{souriau1980groupes,iglesias2013diffeology}.
The key idea is that a higher-order function
is called smooth if it sends smooth functions to smooth functions, meaning that we can never use it
to build non-smooth first-order functions.
This is reminiscent of a logical relation.
\begin{definition}
    A \emph{diffeological space} $X=(\carrier{X},\plots{X})$ consists of a set $\carrier{X}$ together with, for each $n\in\NN$ and each open subset $U$ of $\RR^n$,
     a set $\plots{X}^U$ of functions $U\to\carrier{X}$, called \emph{plots}, such that
	\begin{tightitemize}
	 	\item \textbf{(constant)} all constant functions are plots;
	 	\item \textbf{(rearrangement)} if $f:V\to U$ is a smooth function and $p\in\plots{X}^U$, then $f;p\in\plots{X}^V$;
     \item \textbf{(gluing)} if $\seq[i\in I]{p_i\in\plots{X}^{U_i}}$ is a compatible family of plots $(x\in U_i\cap U_j\Rightarrow p_i(x)=p_j(x))$
     and $\seq[i\in I]{U_i}$ covers $U$,
     then the gluing $p:U\to \carrier{X}:x\in U_i\mapsto p_i(x)$ is a plot.
	 \end{tightitemize} 
\end{definition}
We call a function $f:X\to Y$ between diffeological spaces \emph{smooth} if, for all plots
$p\in\plots{X}^U$, we have that $p;f\in \plots{Y}^U$. We write $\Diff(X,Y)$ for the set of smooth maps from $X$ to $Y$. 
Smooth functions compose, and so we have a category $\Diff$ of diffeological spaces and smooth functions.
We give some examples of such spaces.

\begin{example}[Manifold diffeology]
  Given any open subset $X$ of a cartesian space $\RR^n$ (or, more generally, a smooth manifold $X$).
  We can take the set of smooth $(C^\infty)$ functions $U\to X$ in the traditional sense as $\plots{X}^U$.
  Given another such space $X'$, then $\Diff(X,X')$ coincides precisely with the
  set of smooth functions $X\to X'$ in the traditional sense of calculus and differential geometry.
\end{example}
Put differently, the categories $\CartSp$  of cartesian spaces and $\Man$  of smooth manifolds with smooth functions
form full subcategories of $\Diff$.

\begin{example}[Subspace diffeology]Given a diffeological space $X$ and a 
  subset $|Y|\subseteq |X|$, we can equip it with the \emph{subspace diffeology}
  $\plots{Y}^U\defeq \set{\alpha:U\to|Y|\mid \alpha\in\plots{X}^U}$.
\end{example}
\begin{example}[Coproduct diffeology]
  Given diffeological spaces  $\seq[i\in I]{X_i}$, we can give 
  $\coprod_{i\in I}|X_i|$ the \emph{coproduct diffeology}:
  $\plots{\coprod_{i\in I}X_i}^U\defeq \set{U\xto{\alpha} \coprod_{i\in I}|X_i|\mid \forall i\in I.
  V_i\defeq \alpha^{-1}|X_i|\subseteq_{\mathrm{open}}U\textnormal{ and }
  \alpha|_{V_i}\in \plots{X_i}^{V_i}}$. 
\end{example}

\begin{example}[Product diffeology]
  Given diffeological spaces  $\seq[i\in I]{X_i}$, we can equip 
  $\prod_{i\in I}|X_i|$ with the \emph{product diffeology}:
  $\plots{\prod_{i\in I}X_i}^U\defeq \set{\seq[i\in I]{\alpha_i}\mid \alpha_i\in \plots{X_i}^U}$. 
\end{example}

\begin{example}[Functional diffeology]
  Given diffeological spaces $X,Y$, we can equip $\Diff(X,Y)$ with the
  \emph{functional diffeology} $\plots{Y^X}^U\defeq \set{\Lambda(\alpha)\mid \alpha\in \Diff(U\times X, Y)}$.
\end{example}
These examples give us the categorical (co)product and exponential objects, respectively, in $\Diff$.
The embeddings of $\CartSp$ and $\Man$ into $\Diff$ preserve
products and coproducts.
Further, $\Diff$ is, in fact, complete and cocomplete (and even is a Grothendieck quasi-topos \cite{baez2011convenient}).

Finally, it is well-known that the Euclidean topology that 
is induced on manifolds by their atlas has a canonical extension 
to diffeological spaces, named the \emph{D-topology} \cite{christensen2014d}.
It is defined by
$\Opens X=\set{U\subseteq |X|\mid \forall n\in\NN,V\subseteq_{\mathrm{open}}\RR^n,
\alpha\in\plots{X}^V. \alpha^{-1}U\subseteq_{\mathrm{open}}V}$.
This D-topology defines an (identity on morphisms) adjunction $\Opens{\_} \dashv R:\CTop\to\Diff$ from the category $\CTop$
of topological spaces and continuous functions to diffeological spaces,
where $|RS|\defeq|S|$ and $\plots{RS}^U\defeq \CTop(U,S)$.

\subsection{Partiality in $\Diff$}
The following is new.
We have an obvious candidate $\sem{\reals}\defeq\RR$
which trivially interprets all real constants.
In particular, we can interpret our full language in $\Diff$
if we give a suitable strong monad $\Lift{(-)}$ such that we 
can find suitable definitions for $\sIt$, $\sSign$, and 
$\sem{\op}$.

\subsubsection*{Smooth partial functions}
Ultimately, we are interested in interpreting the first-order 
fragment of our language in the category $\pMap{\Man}$ of 
manifolds and partial functions defined on an open domain 
on which they are smooth.
We are therefore looking for a higher-order extension of 
$\pMap{\Man}$.
As the D-topology simply reduces to the Euclidean topology
on manifolds, it is natural to define a category $\pMap\Diff$
of diffeological spaces and partial functions which restrict to
a smooth function of diffeological spaces on their domain
(equipped with the subspace diffeology), which 
is D-open in $X$.
Then, $\pMap\Man(M, N)\cong \pMap\Diff(M, N)$
for any two manifolds $M,N$,
so that $\pMap\Diff$ can be seen 
as a higher-order extension of $\pMap{\Man}$.

\subsubsection*{A partiality monad}
On the other hand, we have a natural partiality monad on $\Diff$.
\begin{definition}Given a diffeological space $X$, we 
  equip the set $|X|+ \set{\bot}$ with the \emph{partiality diffeology}: 
$
\plots{\Lift{X}}^U\defeq \set{\alpha:U\to |X|+ \set{\bot}\mid V\defeq\alpha^{-1}|X|\subseteq_{\mathrm{open}}U\textnormal{ and } \alpha|_{V}\in\plots{X}^V}.
$
\end{definition}
This definition is easily seen to give a well-defined diffeology.
Note that it is qualitatively different from the coproduct $X+\terminal$ in $\Diff$, as $\alpha^{-1}\set{\bot}$ need not be open.

\begin{proposition}
The partiality monad structure $(\return,\bind)$ from $\Set$ lifts 
to $\Diff$ to make $\Lift{(-)}$ into a commutative strong monad.
\end{proposition}
To be explicit, we define $\return(x)\defeq x$, 
$\bot\bind f \defeq \bot$ and $x\bind f\defeq f(x)$ (for $x\neq \bot$).

It turns out that both natural notions of partiality coincide.
\begin{proposition}
We have an equivalence of categories between $\pMap\Diff$ and the Kleisli 
category $\Lift{\Diff}$ of $\Lift{(-)}$.
\end{proposition}
Indeed, D-open subsets $U\subseteq X$ are precisely those that 
arise as $\chi_U^{-1}(\star)$ for a smooth characteristic function $X\xto{\chi_U}\Lift{\terminal}$. 
In particular, $\Lift{\Diff}(M,N)\cong \pMap\Man(M, N)$ for any two manifolds $M,N$.

\subsubsection*{Interpreting operations}
We have designed our language such that we have a smooth partial function 
$\sem{\op}\in \pMap\Man(\RR^n,\RR)$ in mind for any operation symbol 
$\op\in\Op_n$.
As a result, the interpretation of each operation is fixed.
Later, we will see that, to establish correctness of our AD macro,
we need to have chosen each $\partial_i\op(\var_1,\ldots,\var_n)$ such that 
$\sem{\partial_i\op(\var_1,\ldots,\var_n)}=\nabla_i\sem{\op}$, where we  write $\nabla_i$ for the 
usual calculus partial derivative along the $i$-th coordinate on $\RR^n$.

Next, observe that we have an obvious choice of $\sSign\in\pMap\Man(\RR,\terminal+\terminal)$.
Indeed, we choose the smooth partial function, defined on $\RR\setminus\set{0}$,
which sends $\RR^+$ to $\injection_1\star$ and $\RR^-$ to $\injection_2\star$.

\subsubsection*{Interpreting iteration}
Observe that we can glue matching families of morphisms in 
$\Diff$, by axiom (gluing) of a diffeology, and that the union of any collection of D-opens is 
D-open, by the axioms of a topology.
As a consequence, lubs of compatible families of morphisms in $\pMap\Diff$ exist,
where we call two partial functions compatible if they agree on the intersection
of their domains.
\begin{proposition}
Any compatible family $\seq[i\in I]{f_i}$ of smooth partial functions 
$f_i\in \pMap\Diff(X,Y)$ has a least upper bound in  $\pMap\Diff(X,Y)$:
$$(\lub_{i\in I}f_i)(x)\defeq\left\{\begin{array}{ll} f_i(x) & \exists i\in I. f_i(x)\in Y\\
\bot & else.\end{array}\right.$$
Moreover, composition in $\pMap\Diff$ respects these lubs:
$(\lub_{i\in I}f_i);(\lub_{j\in J}g_j)=\lub_{i\in I}\lub_{j\in J}f_i;g_j$.
Homsets have a least element, the partial function with domain $\emptyset$, and composition is strict in both arguments. 
\end{proposition}
\noindent In particular, $\pMap\Diff\cong \Lift{\Diff}$ (whose equivalence respects the natural orders)
is $\Lift{\wCpo}$-enriched.

\begin{lemma}
  The monad $\Lift{(-)}$ on $\Diff$ has the properties that
  \begin{itemize}
  \item the strength is $\omega$-continuous: $(\id \times \lub_i f_i);\strength =
  \lub_i((\id \times f_i);\strength)$;
  \item copairing in $\Diff_\bot$ is $\omega$-continuous in both arguments: $[\lub_i f_i,\lub_i g_i] = \lub_i[f_i, g_i]$;
  \item the strength and by postcomposition in $\Diff_\bot$ are strict: $(\id \times \bot);\strength = \bot$\textit{ and }
  $\bot\klcomp f = \bot$,
  \end{itemize}
  where we write $\strength$ for the strength $\strength(a,b)=b\bind (\beta\mapsto \sPair a {\return(\beta)})$.
  \end{lemma}
  Indeed, all of these facts follow from the corresponding facts for the category of sets and partial functions
  once we observe that we have a forgetful functor $\Diff\to \Set$ which sends $\Lift{(-)}$ to the usual (strong) partiality
  monad.
  
  Given $f\in \Diff_\bot(A,A+B)$, we can observe that the map
  $
  \Diff_\bot( A,B)\to \Diff_\bot( A,B);\;
  h\mapsto f\klcomp [h,\return[B]]
$
  is $\omega$-continuous.
  As a consequence, we can define $\iterate{f}$ as the least fixpoint of this map.
  As shown in Thm. 5.8 of \cite{goncharov2015unguarded}, this determines a complete Elgot monad structure for $\Lift{(-)}$.
  Therefore, we can interpret iteration constructs $\sIt_{C,A,B}:\Diff_\bot(C\times A,A+B)\to\Diff_\bot(C\times A, B)
  $
  which are natural in $C$.
  Indeed, $\sIt_{C,A,B}(f)$ gets interpreted as the least fixed-point of
  the $\omega$-continuous map $\Diff_\bot(C\times A,B)\to \Diff_\bot(C\times A,B);\;
      h\mapsto (\diag_C\times \id[A]);\assoc[C,C,A];(\id[C]\times f);\strength_{C,A+B}\klcomp(\dist_{C,A,B}; [h,\projection_2;\return[B]]),$
  where $\diag$ is the diagonal of the cartesian structure, $\assoc$ is the associator of $\times$, and $\dist$
  is the distributor of $\times$ over $+$.
  
In sum, we obtain a canonical denotational semantics $\sem{-}$ of our full language, interpreting 
values $\Ginfv \val\ty$ as morphisms $\Diff(\sem{\Gamma},\sem{\ty})$ and computations 
$\Ginfc\trm\ty$ as morphisms $\pMap\Diff(\sem{\Gamma},\sem{\ty})\cong \Lift{\Diff}(\sem{\Gamma},\sem{\ty})$.
In particular, any computation $\Gamma\vdash^c \trm:\ty$ between first-order types is interpreted as a 
smooth partial function $\pMap\Man(\sem{\Gamma},\sem{\ty})$ in the usual calculus sense,
as $\pMap\Man(M,N)\cong \pMap\Diff(M,N)$ for manifolds $M,N$.
\section{Logical Relations and Correct AD for Iteration}\label{sec:logical-relations1}
Using the same strategy as \cite{hsv-fossacs2020}, we prove the
correctness of AD by employing a logical relations proof over the semantics.
We interpret each type $\ty$ as the pair $(\sem{\ty},\sem{\Dsyn{\ty}})$
of diffeological spaces, together with a suitable binary relation $R$ 
relating curves in $\sem{\ty}$ to their tangent curves in $\sem{\Dsyn{\ty}}$.
Such proofs can be cleanly stated in terms 
of subsconing, which we recall now.

\subsection{Preliminaries}
\subsubsection*{Artin glueing/sconing}
Given a functor $F:\catC\to\catD$, the comma category $\catD/F$ is also known as the \emph{scone}
or \emph{Artin glueing} of $F$.
Explicitly, it has objects $(D,f,C)$ which are triples of an object $D$ of $\catD$, an object $C$ of $\catC$ 
and a morphism $f:D\to FC$ of $\catD$.
Its morphisms $(D,f,C)\to (D',f',C')$  are pairs $(g, h)$ of morphisms such that $f;F(h)=g;f'$.
Under certain conditions, $\catD/F$ inherits much of the structure present in $\catC$ and $\catD$, 
letting us build new categorical models from existing ones.
Indeed, $\catD/F$ lifts any coproducts that exist in $\catC$ and $\catD$ \cite{rydeheard1988computational}.
Moreover, if $\catC$ and $\catD$ are cartesian closed, $\catD$ has pullbacks, and $F$ preserves finite products,
$\catD/F$ is cartesian closed \cite{carboni1995connected,johnstone-lack-sobocinski}.
Note that we have canonical projection functors $\catD/F\to\catC$ and $ \catD/F\to\catD$.
Those preserve the bicartesian closed structure.

\subsubsection*{Factorization systems}
Recall that a \emph{orthogonal factorisation system} on a category $\cat$ is a pair
$\pair{\catE}{\catM}$ consisting of two classes of morphisms of
$\cat$ such that:
  \begin{itemize}
  \item
    Both $\mathcal{E}$ and $\catM$ are closed under composition, and contain
    all isomorphisms.
  \item
    Every morphism $f : X \to Y$ in $\cat$ factors into $f = e;m$
    for some $m \in \catM$ and $e \in \mathcal{E}$.
  \item
    Functoriality: for each situation as on the
    left, there is a unique $h : A\to A'$ as on the right:
    \centering
      \begin{tikzcd}[row sep=scriptsize]
        X \arrow[r, "e \in \mathcal E\ \ " above, two heads]
        \arrow[d, "f" left] &
        A \arrow[r, "m \in \mathcal M" above, tail]
        \arrow[d, "=", phantom]
        &
        Y
        \arrow[d, "g" right]
        \\
        X' \arrow[r, "e' \in \mathcal E\ \ " below, two heads] &
        A' \arrow[r, "m' \in \mathcal M" below, tail]
        &
        Y'
      \end{tikzcd}
      $\implies$
      \begin{tikzcd}[row sep=scriptsize]
        X \arrow[r, "e" above, two heads]
        \arrow[d, "f" left]
        \arrow[rd, "=", phantom] &
        A \arrow[r, "m" above, tail]
        \arrow[d, "{h}" , dashed]
        &
        Y
        \arrow[d, "g" right]
        \\
        X' \arrow[r, "e'" below, two heads] &
        A' \arrow[r, "m'" below, tail]
        &
        Y'
        \arrow[ul, "=", phantom] &
      \end{tikzcd}
\end{itemize}
In case $\cat$ is $\wCpo$-enriched, we call  
$(\catE,\catM)$ an \emph{$\wCpo$-enriched factorization system} in case the assignment $(f,g)\mapsto h$ in the 
diagram above is $\omega$-continuous.

\subsubsection*{Subsconing}
Suppose that $\catD$ is equipped with some orthogonal factorization system $(\Cat{E},\Cat{M})$
and that we are given a functor $F:\catC\to\catD$.
We call the full subcategory $\catD//F$ of $\catD/F$ on the objects $(D,f,C)$ where $f\in \catM$ 
the \emph{subscone} of $F$.
Often, $\catM$ will consist exclusively of monos, in which case we can think of the objects of $\catD//F$ 
as objects $C$ of $\catC$ together with a predicate $m:D\monomo FC$ (in $\catM$) on $FC$.
Then, morphisms are simply morphisms $h$ in $\catC$ such that $Fh$ respects the predicate.
$\catD//F$ is easily seen to be a reflective subcategory of $\catD/F$, hence its products are 
computed as in $\catD/F$ and it inherits the existence of coproducts from $\catD/F$. If $\catM$ is closed under 
$I$-ary coproducts, $I$-ary coproducts are computed in $\catD//F$ as in $\catD/F$.
Finally, if $\catM$ is closed under exponentiation (or, equivalently, $\catE$ is closed under binary products),
the exponentiatials of $\catD/F$ also give exponentials in $\catD//F$, making $\catD//F$ cartesian closed.

In case we specialize to the case of $\catD=\Set$ and $F=\catC(\terminal,-)$ and we work with the usual 
epi-mono factorization system, we see that the bicartesian closed 
structure of $\catD//F$ precisely yields the usual formulas of logical predicates over $\catC$ \cite{mitchell1992notes}.
We can thus see (the interpretation of a language) in a subscone as a generalized logical relations argument.

For $\wCpo$-enriched categories $\catC$ and $\catD$ and a locally continuous functor $F:\catC\to\catD$,
$\catD//F$ is a reflective subcategory of $\catD/F$ in the $\wCpo$-enriched sense in case 
$(\catE,\catM)$ is an $\wCpo$-enriched factorization system.

\subsubsection*{Sheaves}
Given a topological space $X=(|X|,\Opens X)$, a presheaf on the open subsets
$\Opens X$ is a contravariant functor $F$ from the poset $\Opens X$ to $\Set$.
Given an open cover $\cover$ of $W\in\Opens X$,
a matching family of $F$ for $\cover$ is a tuple $\seq[U\in\cover]{x_U\in F(U)}$ such that 
$F(V\subseteq U)(x_U)=F(V\subseteq U')(x_{U'})$ for all $U,U'\in\cover$ and $V\subseteq_{\mathrm{open}}U\cap U'$.
An amalgamation of such a matching family is an element $x\in FW$ such that $F(U\subseteq W)(x)=x_U$,
for all $U\in\cover$.
A presheaf $F$ on $\Opens X$ is called a sheaf if every matching family has a unique amalgamation.
Sheaves and natural transformations form a cartesian closed category $\Sh(\Opens X)$ with all small limits and colimits.
(In fact, they form a Grothendieck topos.)
Further, epis and monos (i.e. componentwise injective natural transformations) form an orthogonal factorization system on $\Sh(\Opens X)$.
By a subsheaf $G$ of a sheaf $F$, we mean a sheaf $G$ such that $GU\subseteq FU$ for all $U\in\Opens X$
and $G(U\subseteq U')$ is the restriction of  $F(U\subseteq U')$ for all $U\subseteq U'\in \Opens X$.

\subsection{Subsconing for correctness of AD}
In \cite{hsv-fossacs2020}, the idea is to build an interpretation $\semgl{-}$ of the language in the subscone 
$\Diff\times \Diff//\Diff\times\Diff((\RR,\RR),-)$, where we interpret 
each $\ty$ as the pair $(\sem{\ty},\sem{\Dsyn{\ty}})$ together with the relation 
which relates each curve in $\sem{\ty}$ to its tangent curve in $\sem{\Dsyn{\ty}}$.
This category almost has enough structure to lift the interpretation of our language:
it is bicartesian closed. All that is missing is a lifting of the partiality monad $\Lift{(-)}$.
One may try to define a lifting $\RLift{(-)}$ sending relations $R$ on $(\RR\to X)\times (\RR\to Y)$
to relations $(\RR\to \Lift{X})\times (\RR\to\Lift{Y})$.
A natural definition states that $\RLift{R}$ contains those pairs $(\gamma,\gamma')$ of 
partially defined curves in $X$ and $Y$, such that they have the same domain and are in $R$ when 
restricted to their domain.
Indeed, to prove AD correctness, we want the relation $R$ of $\semgl{\reals}$ to say that
the second curve has the same domain as the first and is the derivative of the first on its domain.
However, a priori, we cannot make sense of the statement that $(\gamma,\gamma')\in R$ on their domain,
as $R$ only tracks curves with domain $\RR$!
Therefore, we need to generalize to relations which track arbitrary partially defined curves.
It turns out that it is best to treat such relations on partially defined curves as sheaves
of relations over $\Opens\RR$
(rather than as mere tuples or presheaves of relations)
if we want the proof apply to sum types.

Thus, using the fact that one-dimensional plots form a sheaf over $\Opens \RR$
by the (rearrangement) and (gluing)
axioms, we consider the subscone $\Gl $ of the functor
\begin{align*}
    F_{\Gl }:\Diff\times\Diff&\to\Sh(\Opens\RR)\qquad
    (X,Y)\mapsto \plots{X}\times \plots{Y},
\end{align*}
where we work with the epi-mono factorization system on $\Sh(\Opens \RR)$.
Concretely, objects of $\Gl $ are triples $(X,Y,R)$ of two diffeological spaces 
$X,Y$ and a subsheaf $R$ of $\plots{X}\times\plots{Y}$.
We unroll the latter further: $R$ gives for every $U\subseteq_{\mathrm{open}}\RR$ 
a relation $R^U\subseteq \plots{X}^U\times \plots{Y}^U$ such that relations restrict and 
glue in the sense that 
\begin{itemize}
\item if $V\subseteq_{\mathrm{open}} U$ and $(\gamma,\gamma')\in R^U$, then 
$(\gamma|_V,\gamma'|_V)\in R^V$;
\item if $\gamma\in\plots{X}^W$ and $\gamma\in\plots{Y}^W$ and $\cover$ is an open cover of 
$W$ such that $(\gamma|_U,\gamma'|_U)\in R^U$ for all $U\in\cover$, then also $(\gamma,\gamma')\in R^W$.
\end{itemize}
Morphisms $(X,Y,R)\to(X',Y',R')$ in $\Gl $ are pairs $(X\xto{g}X',Y\xto{h}Y')$ of diffeological space morphisms 
such that, for any $U\in\Opens\RR$, $(\gamma,\gamma')\in R^U$ implies that $(\gamma;g,\gamma';h)\in R'{}^U$.
By the general theory for subsconing discussed above, $\Gl $ lifts the bicartesian closed structure of $\Diff\times \Diff$ 
along the obvious forgetful functor $\Gl \to\Diff\times\Diff$.
To lift $\Lift{(-)}\times\Lift{(-)}$ to $\Gl $, we define,  given $(X,Y,R)$,
$$\RLift{R}^U\defeq \set{\sPair{\gamma}{\gamma'}\in\plots{\Lift{X}}^U\times\plots{\Lift{Y}}^U\mid \gamma^{-1}(X)=\gamma'{}^{-1}(Y)\textnormal{ and }
\sPair{\gamma|_{ \gamma^{-1}(X)}}{\gamma'|_{ \gamma^{-1}(X)}}\in R^{{ \gamma^{-1}(X)}}}.$$
This definition is easily seen to give a subsheaf of $\plots{\Lift{X}}\times\plots{\Lift{Y}}$.
\begin{proposition}
The strong monad operations of $\Lift{(-)}\times\Lift{(-)}$ lift to $\Gl $, turning 
$\RLift{(-)}$ into a commutative strong monad.
\end{proposition}
One can either verify this proposition by hand or observe that it follows from the general theory of logical relations for monadic types of \cite{goubault2002logical},
if we work with the partiality monad ($(TF)U\defeq \set{(V,x)\mid V\in\Opens U\textnormal{ and } x\in FV}$,
$\return[U](x)\defeq (U,x)$, $(V,x)\bind_U \alpha_U\defeq \alpha_V(x)$) on $\Sh(\Opens \RR)$
together with the obvious distributive law $F_{\Gl };T\to \Lift{(-)};F_{\Gl }$.

\begin{corollary}
    The lifted monad $\RLift{(-)}$ lifts the interpretation of iteration in $\Lift{(-)}\times\Lift{(-)}$.
    \end{corollary}
    \begin{proof}
    Indeed, the interpretation of iteration is constructed entirely out of the structure of the strong monad, the bicartesian closed structure,
    and the $\pMap{\wCpo}$-enrichment of $\Lift{\Diff}\times\Lift{\Diff}$.
    Seeing that $\RLift{\Gl }$ lifts all of this structure, it lifts the interpretation of iteration as well.
\end{proof}
    
\subsection{Correctness of Forward AD}\label{ssec:correctness-of-forward-ad}
Next, we build an interpretation $\semgl{-}$ of our language in $\Gl $
lifting the interpretation $\sPair{\sem{-}}{\sem{\Dsyn{-}}}$ of our language 
in $\Diff\times\Diff$
and show that correctness of forward AD follows.
If we unwind all the definitions of the categorical structure of $\Gl $, the subsconing argument
can be seen as the 
elementary logical relations proof described in section \ref{sec:key-ideas}.

Suppose we are given a smooth partial function $f:U\rightharpoonup{V}$
between open subsets $U\subseteq \RR^n$ and $V\subseteq \RR^m$
and a vector $v\in \RR^n$.
We write $\nabla_v f$ for the smooth partial function $U\rightharpoonup\RR^m$
which has the same (open) domain $D$ as $f$ and calculates the
directional derivative of $f|_D$ in direction $v$ on $D$.
We will slightly abuse notation and write $\nabla_i f$ for some $1\leq i\leq n$ to mean $\nabla_{e_i}f$ where $e_i$ is the $i$-th 
standard basis vector of $\RR^n$.
Similarly, in case $U$ is one-dimensional, we write $\nabla f$ for $\nabla_{e_1}f$.

To establish correctness of $\Dsyn{-}$ with respect to the semantics $\sem{-}$, it is important to note that
$$
\semgl{\reals}^V\defeq \set{\sPair{f}{\sPair{f}{\nabla f}}\mid f\in \plots{\RR}^V}\subseteq \plots{\RR}^V\times \plots{\RR\times\RR}^V
$$
forms a sheaf on $\Opens \RR$, hence an object in $\Gl $.
It forms a sheaf because differentiation is a local operation,
in the sense that (1) derivatives restrict to open subsets $U$ -- $\left[\nabla (f|_U)=(\nabla f)_U\right]$
and (2) derivatives glue along open covers $\cover$ -- $\left[\nabla\glue\seq[U\in\cover]{f_U}=
\glue\seq[U\in\cover]{\nabla f_U}\right]$.

Observe that $\sPair{\sem{\tSign}}{\sem{\Dsyn{\tSign}}}=\sPair{\sSign}{(\pi_1;\sSign)} \in \Gl (\semgl{\reals},\RLift{\semgl{\Unit\t+\Unit}})$:
Indeed, assume that $\sPair{\alpha}{\alpha'}\in \semgl{\reals}^U$.
That is, $\alpha'=\sPair{\alpha}{\nabla \alpha}$.
We show that
$\sPair{\alpha;\sSign}{\alpha;\sSign}=\linebreak \sPair{\alpha}{\alpha'};\sPair{\sem{\tSign}}{\sem{\Dsyn{\tSign}}}\in
\RLift{\semgl{\Unit\t+\Unit}}^U$:
using the definition of the coproduct in $\Gl $, we have that
\begin{align*}
    \RLift{\semgl{\Unit\t+\Unit}}^U &= \set{\sPair{\beta}{\beta'} \mid
    \beta^{-1}(\terminal+\terminal) = \beta'^{-1}(\terminal+\terminal)\textnormal{ and } \sPair{\beta|_{\beta^{-1}(\terminal+\terminal)}}{\beta'|_{\beta'^{-1}(\terminal+\terminal)}}\in \semgl{\Unit\t+\Unit}
    }\\
    &= \set{\sPair{\beta}{\beta'}   \mid
    \beta^{-1}(\terminal+\terminal) = \beta'^{-1}(\terminal+\terminal)\textnormal{ and } {\beta|_{\beta^{-1}(\terminal+\terminal)}}={\beta'|_{\beta'^{-1}(\terminal+\terminal)}}
    }\\
    &=\set{\sPair{\beta}{\beta}\mid \beta \in \Diff(U,\Lift{(\terminal+\terminal)})}.
    \end{align*}
For these equalities to hold, it is important that we work with sheaves and not mere presheaves.

Then, as long as the operations respect the relation,
which follows if $\sem{\partial_i\op(\var_1,\ldots,\var_n)}=\nabla_i\sem{\op(\var_1,\ldots,\var_n)}$, for all $n$-ary 
operations $\op$ and all $1\leq i \leq n$,
it follows immediately that $\semgl{-}$ extends uniquely to the full language to lift the interpretation $\sem{\sem{-}}{\sem{\Dsyn{-}}}$.
(This equation is precisely the criterion that $\partial_i\op$ is a correct implementation of the $i$-th partial derivative of $\op$.)
Indeed, observe that $\semgl{-}$ and the projection functor $\projf:\Gl \to \Diff\times\Diff$
are structure preserving, as are $(\id,\Dsyn{-})$ and $\sem{-}\times\sem{-}$.
Meanwhile, $(\id,\Dsyn{-});\sem{-}\times \sem{-}$ and $\semgl{-};\projf$ agree on $\reals$ and all operations $\op$ and $\tSign$.
By the universal property of the syntax, therefore, $\semgl{-};\projf=\sPair{\sem{-}}{\sem{\Dsyn{-}}}$.

In particular, for any computation $\var_1:\reals,\ldots,\var_n:\reals\vdash^c \trm:\reals$, we have that 
$(\sem{\trm},\sem{\Dsyn{\trm}})\in\RLift{(\Gl )}(\semgl{\reals}^n,\semgl{\reals})$.
Given a point $x\in\RR^n$ and a tangent vector $v\in \RR^n$, there exists some smooth curve 
$\gamma:\RR\to \RR^n$ such that $\gamma(0)=x$ and $\nabla\gamma(0)=v$. 
Now, by the chain rule, $(\gamma, (\gamma,\nabla\gamma))\in \Gl (\semgl{\reals},\semgl{\reals}^n)$.
Therefore, $(\gamma;\sem{\trm}, (\gamma,\nabla\gamma);\sem{\Dsyn{\trm}})\in \Gl (\semgl{\reals},\semgl{\reals})$.
That is,
$
(\gamma,\nabla\gamma);\sem{\Dsyn{\trm}} =(\gamma;\sem{\trm}, \nabla (\gamma;\sem{\trm}))
$
and so, in particular, evaluating at $0$,
$\sem{\Dsyn{\trm}}(x,v)$ is defined iff $\sem{\trm}(x)$ is defined and in that case
$
\sem{\Dsyn{\trm}}(x,v)=(\sem{\trm}(x),\nabla_v\sem{\trm}(x)).
$
\begin{corollary}[Semantic Correctness of $\Dsynsymbol$ (limited)]\label{cor:sem-cor-limited}
    For any $\var_1:\reals,\ldots,\var_n:\reals\vdash^c\trm :\reals$,
    we have that $\sem{\DsynC{\trm}}(x, v)=(\sem{\trm}(x), \nabla_{x}\sem{\trm}v)$, for all $x$ in the domain of 
    $\sem{\trm}$ and $v$ tangent vectors at $x$.
    Moreover, $\sem{\DsynC{\trm}}(x,v)$ is defined iff $\sem{\trm}(x)$ is.
\end{corollary}
\section{Correctness of forward AD at first-order types} \label{sec:general-correctness1}
We can extend this correctness result to functions between arbitrary first-order types.
\subsection{Preliminaries}
\subsubsection*{Tangent bundles}

We recall that the derivative of any smooth function $f:M\to N$ between manifolds 
is given as follows.
For each point $x$ in a manifold $M$, define the \emph{tangent space} $\Dsemsymbol_x M$ to be the set $\{\gamma\in\Man(\RR,M)\mid \gamma(0)=x\}/\sim$ of equivalence classes $[\gamma]$ of smooth curves $\gamma$
in $M$ based at $x$, where we identify $\gamma_1\sim \gamma_2$ iff $\nabla (\gamma_1;f)(0)=\nabla(\gamma_2;f)(0)$ for all smooth $f:M\to\RR$.
The \emph{tangent bundle} of $M$ is the set $\Dsem{M}\defeq \biguplus_{x\in M} \Dsemsymbol_x (M)$. The charts of $M$ equip $\Dsem{M}$ with a canonical manifold structure.
Then for smooth $f:M\to N$, the derivative $\Dsem{f}:\Dsem{M}\to\Dsem{N}$ is defined as $\Dsem{f}\sPair{x}{[\gamma]}\defeq \sPair{f(x)}{[\gamma;f]}$.
By the chain-rule, $\Dsemsymbol$ is a functor $\Man\to\Man$. 
As is well-known, we can understand the tangent bundle of a composite space in terms of that of its parts.
\begin{lemma}[Tangent Bundles of (Co)Products]\label{lemma:Dsem}
For countable $I$ and $n\in\NN$, there are canonical isomorphisms
$\Dsem{\biguplus_{i\in I} M_i} \cong \biguplus_{i\in I} \Dsem{M_i}$ and 
$\Dsem{M_1\times\ldots \times M_n}\cong \Dsem{M_1}\times\ldots\times \Dsem{M_n}$.
\end{lemma}

\subsection{Correctness at first-order types}
We define a canonical isomorphism $\DtoT{\ty}:\sem{\Dsyn{\ty}}\to\Dsem{\sem{\ty}}$ for every type $\ty$, by induction on the structure of types. We let $\DtoT{\reals}:\sem{\Dsyn{\reals}}\to\Dsem{\sem{\reals}}$ be given by $\DtoT{\reals}(x,x')\defeq (x,[t\mapsto x+x't])$.
For the other types, we use Lem.~\ref{lemma:Dsem}. 

Next, we note that the tangent bundle functor $\Dsemsymbol:\Man\to \Man$ extends to a
functor $\Dsemsymbol$ on the category of manifolds and partial maps $\pMap\Man\to\pMap\Man$.
Indeed, given a smooth partial function $f:M\rightharpoonup{N}$ between manifolds, we define
$\Dsem{f}:\Dsem{M}\rightharpoonup{\Dsem{N}}$ to have domain $\Dsem{f^{-1}(N)}\subseteq \Dsem{M}$ and definition 
$\Dsem{f}|_{\Dsem{f^{-1}(N)}}\defeq\Dsem{f|_{f^{-1}(N)}}$.

This extension lets us extend the correctness result of Cor. \ref{cor:sem-cor-limited}
to functions between arbitrary first-order types, i.e. types which do not contain any 
function type constructors.
 \begin{theorem}[Semantic correctness of $\Dsynsymbol$ (full)]\label{thm:fwd-cor-full}
   For any first-order type $\ty$, any first-order context $\Gamma$
   and any  computation $\Gamma\vdash^c\trm:\ty$,
   the syntactic translation $\Dsynsymbol$ coincides with the tangent bundle functor, modulo these canonical isomorphisms:
   $\sem{\DsynC{\trm}};\DtoT{\ty}=\DtoT{\ty};\Dsem{\sem{\trm}}$.
 \end{theorem}
 \begin{proof}
    For any partial curve $\gamma\in\pMap\Man(\RR,M)$, let $\bar \gamma\in\pMap\Man(\RR,\Dsem M)$ be the tangent curve, given by $\bar \gamma(x)=(\gamma(x),[t\mapsto \gamma(x+t)])$,
    which has the same domain as $\gamma$.

    First, we note that a smooth partial map map $h\in\pMap\Man(\Dsem M, \Dsem N)$ is of the form $\Dsem{g}$ for some $g\in\pMap\Man(M,N)$
    if for all smooth partial curves $\gamma\in\pMap\Man(\RR, M)$ we have $\bar \gamma;h=\overline{(\gamma;g)}\in\pMap\Man(\RR, \Dsem N)$.
   Therefore, it is enough to show that $\tilde{\gamma};\DtoT{\Gamma};\Dsem{\sem{\trm}} =\tilde\gamma; \sem{\DsynC{\trm}};\DtoT{\tau}$
   for all smooth partial curves $\gamma\in\pMap\Man(\RR,\sem{\Dsyn{\Gamma}})$, where $\tilde\gamma$ is the unique smooth partial curve in $\pMap\Man(\RR, \sem{\Dsyn{\Gamma}})$
   such that $\tilde\gamma;\DtoT{\Gamma}=\bar \gamma$.

Second, for any first-order type $\ty$, $\semgl{\ty}^\RR=\{(f,\tilde{f})~|~f:\RR\to \sem{\ty}\}$.
This equation is shown by induction on the structure of types. 
For this result to hold at sum types, it is important that our logical relations are sheaves (and not mere presheaves).
Moreover, for such $\ty$, $\RLift{{\semgl{\ty}}}^\RR=\{(f,\tilde{f})~|~f:\RR\rightharpoonup{\sem{\ty}}\}$. 

We conclude the theorem by combing these two observations:
\begin{align*}\tilde \gamma;\DtoT{\Gamma};\Dsem{\sem{\trm}}
 =\bar\gamma;\Dsem{\sem{\trm}}
 =\overline{\gamma;\sem{\trm}}
 =\widetilde{\gamma;\sem{\trm}};\DtoT{\ty}
 =\tilde\gamma;\sem{\DsynC{\trm}};\DtoT{\tau},
\end{align*}
where in the last step, we use the fact that $\sPair{\sem{\trm}}{\sem{\DsynC{\trm}}}$ respects the logical relation.
\end{proof}

\section{A Language with Term and Type Recursion}
\label{sec:language2}

We extend our language of Sec. \ref{sec:language1} with term
and type recursion. We work with iso-recursive types.
That is, we add the following
types $\ty,\ty[2],\ty[3]$, values $\val,\val[2],\val[3]$, and computations $\trm,\trm[2],\trm[3]$.\\
\begin{syntax}
  \ty, \ty[2], \ty[3] & \gdefinedby & & \syncat{types}                          \\
  &\gor& \ldots                      & \synname{as before}\\
  &&&\\
  \val, \val[2], \val[3] & \gdefinedby & & \syncat{values}                          \\
  &\gor& \ldots                     & \synname{as before}\\
  &&&\\
  \trm, \trm[2], \trm[3] & \gdefinedby & & \syncat{computations}                          \\
  &\gor & \ldots & \synname{as before}\\
\end{syntax}%
~
\begin{syntax}
  &\gor& \tvar, \tvar[2],\tvar[3]                      & \synname{type variables}\\
  &\gor\quad\, & \rec{\tvar}{\ty} & \synname{recursive types}\\
  &&&\\
  &\gor& \tRoll{\val}                      & \synname{recursive type constructors}\\
  &&&\\
  &&&\\
  &\gor\quad\, & \rMatch{\val}{\var}{\trm} & \synname{recursive type match}\\
\end{syntax}
\\
We extend the type system with the rules of Fig. \ref{fig:types2}.
Every typing judgement should now be read relative to a kinding context
$\Delta=\tvar_1,\ldots,\tvar_n$ which contains the free type variables
used in the judgement, where $\rec{\tvar}{}$ binds the right-most 
type variable in the kinding context.
As this kind-system is standard (see e.g. \cite{vakar2019domain}), we leave it implicit for ease of notation.

\begin{figure}[b]
\framebox{\scalebox{1.0}{\begin{minipage}{\linewidth}\noindent\input{type-system2}\end{minipage}}}
\caption{Typing rules for term and type recursion. As usual, we treat $\rec{\alpha}{}$ as a type level 
variable binder.\label{fig:types2}}
\end{figure}
As is well-known, at this point, iteration and term recursion can be defined as sugar in terms of
the primitives for type recursion (and their typing rules become derivable) \cite{abadi1996syntactic}:
\begin{align*}
&\rec{\var[3]}{\trm}\defeq \letin{body}{\left(\fun{\var}{\fun{\var[2]}{\rMatch{\var}{\var'}{\toin{\var[3]}{(\tReturn \var')\var}{\trm\,\var[2]}}}}\right)}{(\tReturn body)(\tRoll{body})}\\
& \tItFrom{\trm}{\var}{\val}\defeq \left(\rec{\var[3]}{\fun{\var}{\toin{\var[2]}{\trm}{\bvMatch{\var[2]}{\var'}{(\tReturn \var[3])\var'}{\var''}{\tReturn \var''}}}}\right)\, \val.
\end{align*}

We extend the forward AD macro in the unique structure preserving way:
\[
\begin{array}{lll}
\Dsyn{\tvar}\defeq\tvar& \Dsyn{\rec{\tvar}{\ty}}\defeq \rec{\tvar}{\Dsyn{\ty}} &\DsynC{\rMatch{\val}{\var}{\trm}}\defeq \\
\DsynC{\rec{\var}{\trm}}\defeq \rec{\var}{\DsynC{\trm}} & \DsynV{\tRoll{\val}}\defeq \tRoll{\DsynV{\val}} &
\quad\rMatch{\DsynV{\val}}{\var}{\DsynC{\trm}}.
\end{array}    
\]

Lemma \ref{lem:functorial-macro} is easily seen to continue to hold on this extended language,
when we add the $\beta\eta$-rules of Fig. \ref{fig:beta-eta2}.
We can observe that the AD rules for iteration and term recursion can be derived from those for 
type recursion.
\begin{figure}[b]
    \framebox{\scalebox{1.0}{\begin{minipage}{\linewidth}\noindent
 \input{beta-eta2}
 \end{minipage}}}
 \caption{Standard $\beta\eta$-laws for recursive types (which imply rules for term recursion and iteration.). \label{fig:beta-eta2}}
 \end{figure}

\section{$\omega$-diffeological spaces and differentiable semantics of recursion}
\label{sec:semantics2}

\subsection{Preliminaries}

\subsubsection*{Bilimit compact expansions}
We recall some basic machinery for solving recursive domain equations \cite{levy2012call}. Recall that
an embedding-projection-pair (ep-pair) $u : A\epto B$ in an
$\wCpo$-enriched category $\cat[4]$ is a pair consisting of a
$\cat[4]$-morphism $\emb u:A\to B$, the \emph{embedding}, and a
$\cat[4]$-morphism $\prj u:B\to A$, the \emph{projection}, such that
$p;e \leq \id$ and $e;p= \id$. An \emph{embedding} $u
: A \embedding B$ is the embedding part of some ep-pair $A \epto
B$.

An \emph{\wchain{} of ep-pairs} $\pair{\seq[n \in
    \NN]{A_n}}{\seq[n \in \NN]{a_n}}$ in $\cat[4]$
consists of a countable sequence of objects $A_n$ and a countable
sequence of ep-pairs $a_n : A_n \epto A_{n+1}$. A \emph{bilimit}
$\pair Dd$ of such an \wchain{} consists of an object $D$ and a
countable sequence of ep-pairs $\smash{d_n : A_n \epto D}$ such that,
for all $n \in \NN$, $\smash{a_n;d_{n+1} = d_n}$, and
$\smash{\lub_{n \in \NN} \prj d_n; \emb d_n   =
  \id[D]}$. The celebrated \emph{limit-colimit
  coincidence}~\cite{smyth-plotkin:rde} states that the bilimit
structure is equivalent to a colimit structure $\pair{D}{\emb d}$ for
$\pair*{\seq{A_n}}{\seq{\emb a_n}}$, in which case $\prj d_n$ are
uniquely determined, and similarly equivalent to a limit structure
$\pair*{D}{\prj d}$ for $\pair*{\seq{A_n}}{\seq{\prj a_n}}$, in
which case $\emb d_n$ are uniquely determined.

A \emph{zero object} is an object that is both initial and
terminal. A zero object in an $\wCpo$-category~is an \emph{ep-zero object}  if
every morphism into it is a projection and every
morphism out of it is an~embedding.

A \emph{bilimit compact category} is an $\wCpo$-category $\cat[4]$
with an ep-zero and ep-pair \wchain{} bilimits.
When $\cat[3], \cat[4]$ are bilimit compact, every
locally continuous, mixed-variance functor $F :
\opposite{\cat[3]}\times\opposite{\cat[4]}\times{\cat[3]}\times\cat[4]
\to \cat[4]$ has a parameterised solution to the recursive equation
$\semroll : F(A, X, A', X) \xto{\isomorphic} X$, for every $A$, $A'$ in
$\cat[3]$, qua the bilimit $\pair{\mu B.F(A,B,A',B)}{d_{F, A, A'}}$
of
\[
\initial \epto F(A, \initial, A'\!, \initial)
\xepto{\ep F(A, \peto, A'\!, \epto)} F(A, F(A, \initial, A'\!, \initial), A'\!, F(A, \initial, A'\!, \initial)) \epto \cdots
\epto F^n(A, A') \epto \cdots
\]
The solution is minimal in the sense of \citet{pitts1996relational}.  The
assignments $\mu B. F(A, B, A', B)$ extend to a mixed-variance functor
$\mu B. F(-, B, -, B) : \opposite{\cat[3]}\times\cat[3] \to \cat[4]$
by $\mu B. F(f, B, g, B) := \lub_{n}\prj d_n;F^n(f,
g); \emb d_n $.

Finally, a \emph{bilimit compact expansion} $J : \cat \embed \cat[4]$
is a triple consisting of an $\wCpo$-category $\cat$, a bilimit
compact category $\cat[4]$; and an identity-on-objects, locally
continuous, order reflecting functor $J : \cat \to \cat[4]$ such that,
for every ep-pair \wchain{}s $\pair Aa$, $\pair Bb$ in $\cat[4]$,
their bilimits $\pair Dd$, $\pair Ee$, and countable collection of
$\cat$-morphisms $\seq[n \in \NN]{\alpha_n : A_n \to B_n}$ such
that for all $n$:
\[
    \emb a_n;J\alpha_{n+1}  = J\alpha_n;\emb b_n, \qquad \prj a_n; J\alpha_{n} = J\alpha_{n+1};\prj b_n
\]
(i.e., $J\alpha : \pair A{\emb a} \to \pair B{\emb b}$ and $J\alpha
: \pair A{\prj a} \to \pair B{\prj b}$ are natural transformations),
there is a $\cat$-morphism $f : D \to E$ such that $Jf = \lub_{n}
\prj d_n;\alpha_n;\emb e_n$. The motivation for
this definition: given two bilimit compact expansions $I :
\cat[2] \embed \cat[3]$, $J : \cat \embed \cat[4]$, and two
locally continuous functors \[ F : \opposite{\cat[2]}\times \opposite\cat \times \cat[2]
\times \cat \to \cat \quad G : \opposite{\cat[3]}\times
\opposite{\cat[4]} \times \cat[3] \times \cat[4] \to \cat[4]
\quad\text{s.t.}\quad \opposite I\times \opposite J \times I
\times J ; G= F;J
\]
the functor $\mu B. G(-, B, -, B) : \opposite{\cat[3]}\times \cat[3]
\to \cat[4]$ restricts to $\mu B. G(\opposite J-, B, J-, B) :
\opposite{\cat[2]}\times \cat[2] \to \cat$.
Bilimit compact expansions are closed under small products, opposites,
and exponentiation with small categories~\cite{levy2012call}.

\subsubsection*{Semantics of recursion}
We again restrict our attention to monad models as considered
in Sec. \ref{sec:semantics1}.
Suppose that we are given a bicartesian closed category $\catC$
with a strong monad $T$ such that the Kleisli adjunction gives 
rise to a bilimit compact expansion $J:\catC\embed\catC_T$.
Suppose further that the copairing $[-,-]$ and 
the exponential adjunction $-\times A\dashv (-)^A$ 
in $\catC$ are locally continuous.
As the left adjoint $(-)\times A$ preserves all colimits, hence lubs,
it follows that $(-)\times(-)$, $(-)+(-)$ and $(T-)^{(-)}$ extend
to locally continuous bifunctors
$(-)\otimes (-)$, $(-)\oplus (-)$, $(-)\pexp (-)$
on $\catC_T$:
\begin{align*}
f\otimes f' &\defeq (f\times f');\doubstrength\quad
f\oplus f'  \defeq (f + f');[T\sInl, T\sInr]\quad
f\pexp f'  \defeq \Lambda(\sPair{\id}{\Lambda(\sSnd;f');\bind});\return.
\end{align*}
We describe how to extend the interpretation of Sec. 
\ref{sec:semantics1}  to the full language of Sec. 
\ref{sec:language2} in a category with such structure.
We write $\sem{\Delta}_v\defeq \prod_{\alpha\in\Delta}\catC$
and $\sem{\Delta}_c\defeq \prod_{\alpha\in\Delta}\catC_T$ for a kinding context $\Delta$.
We interpret types in context $\Delta$ as locally continuous functors
$\sem{\Delta}_c^{op}\times\sem{\Delta}_c\to \catC_T$
which restrict to $\sem{\Delta}_v^{op}\times\sem{\Delta}_v\to \catC$.
We interpret values and computations as corresponding natural transformations.
We extend $\sem{-}$ of Sec. \ref{sec:semantics1}
to apply componentwise, using $\otimes, \oplus, \pexp$ to interpret
product, sum and function types as locally continuous functors on $\catC_T$, and we
interpret
\begin{align*}
    &\sem{\tvar}\defeq \projection_2;\projection_{\alpha}
    && \sem{\rec{\tvar}\ty}\defeq \rec{X}\sem{\ty}(-,\ldots,-,\alpha,-,\ldots,-,\alpha)\\
    &\sem{\Delta|\Gamma \vdash^v \ty} \defeq \catC^{\sem{\Delta}_v}(\sem{\Gamma}, \sem{\ty}) && \sem{\Delta |\Gamma\vdash^c \ty}\defeq \catC_T^{\sem{\Delta}_c}(\sem{\Gamma},\sem{\ty})\\
    &\sem{\tRoll\val}(\rho) \defeq \semroll(\sem{\val}(\rho)) &&  \sem{\rMatch{\val}{\var}{\trm}}(\rho)\defeq \sem{\trm}(\semroll^{-1}(\sem{\val}(\rho))).
\end{align*}

\subsubsection*{Scott topology}
Given an \wcpo{} $(|X|,\leq_X)$, there is a canonical 
topology on $|X|$ called the \emph{Scott topology}.
A subset $U\subseteq |X|$ is Scott-open iff it is 
arises as $\chi_U^{-1}(\star)$ for some $\omega$-continuous characteristic 
function $X\xto{\chi_U}\Lift{\terminal}$.
In explicit terms, Scott open subsets are precisely those for which 
\begin{itemize}
\item $x\leq y$ and $x\in U$ implies $y\in U$ (upwards closure);
\item $x = \lub_{i\in\NN} x_i$ and $\forall i\in\NN. x_i\not\in U$ implies $x\not\in U$ (\wchain{} inaccessibility).
\end{itemize}
Let us write $\pMap\wCpo$ for the category of \wcpos{} and partial functions 
with a Scott open domain on which they are \wcont{}.
Then, we have an equivalence $\pMap\wCpo\cong\Lift{\wCpo}$.

\subsection{$\omega$-diffeological spaces}
As far as we are aware, no established mathematical 
theory is suitable for describing both differentiation 
and recursion.
For this reason, we introduce a new notion of an 
\emph{$\omega$-diffeological space}, or, briefly, \emph{\wds{}}.
\begin{definition}
An \emph{\wds{}} consists of a triple  $(|X|,\plots{X},\leq_X)$
such that $(|X|,\plots{X})$ is a diffeological space and $(|X|,\leq_X)$ is an \wcpo{}
satisfying the additional condition that
$(\lub_{i\in\NN} \alpha_i)(x)\defeq
\lub_{i\in \NN} (\alpha_i(x))$ 
defines a plot $\lub_{i\in\NN} \alpha_i\in\plots{X}^U$,
for any \wchain{} $\seq[i\in\NN]{\alpha_i\in\plots{X}^U}$ in the pointwise order.
Homomorphisms $X\to Y$ of such structures are defined to be functions 
$f:|X|\to|Y|$, such that $f:(|X|,\plots{X})\to(|Y|,\plots{Y})$ is a 
morphism of diffeological spaces and $f:(|X|,\leq_X)\to(|Y|,\leq_Y)$ is
$\omega$-continuous.
We write $\wDiff$ for the category of \wds es and their 
homomorphisms.
\end{definition}

\begin{example}
Any diffeological space $X$ can be interpreted as an 
\wds{} by taking $\leq_X$ as the 
discrete order $=_X$.
In particular, this is true for any Cartesian space or manifold.
\end{example}

\begin{example}
Let $X$ be an \wds.
Any subset $A\subseteq |X|$ which is an \wcpo{} under the induced order from $X$ 
is an \wds{} under the subspace diffeology.
\end{example}

\begin{example}
Given a family $\seq[i\in I]{X_i}$ of \wds{}es,
we can form their product by equipping $\prod_{i\in I}|X_i|$
with the product diffeology and product order.
\end{example}

\begin{example}
    Given a family $\seq[i\in I]{X_i}$ of \wds{}es,
    we can form their coproduct by equipping $\coprod_{i\in I}|X_i|$
    with the coproduct diffeology and coproduct order.
\end{example}

\begin{example}
    Given \wds{}es $X,Y$, we can equip $\wDiff(X,Y)$ with the pointwise order 
    and
    {diffeology} $\plots{Y^X}^U\defeq \set{\Lambda(\alpha)\mid \alpha\in \wDiff(U\times X, Y)}$.
  \end{example}
These examples give us the categorical (co)product and exponential objects, respectively, in $\wDiff$.
In fact, in Appx. \ref{appx:locally-presentable} we give various characterizations of 
$\wDiff$, which, in particular, show that the category is locally presentable,
hence complete and cocomplete.

Next, we turn to the interpretation of partiality in $\wDiff$.
Let us call subsets D-Scott open if they are both D-open and Scott open.
There is a natural notion of partial function $X\rightharpoonup Y$ in $\wDiff$:
partial functions $|X|\rightharpoonup |Y|$ which are defined on a D-Scott open subset $D\subseteq |X|$ 
on which they restrict to an \wds{} morphism $D\to Y$ (using the subspace order and diffeology).
Let us write $\pMap\wDiff$ for the category of \wds{}es and such partial functions.
As we interpret first-order types as manifolds with the discrete order in $\wDiff$, 
the semantics of computations between first-order types again lies in the full subcategory $\pMap\Man$ of $\pMap\wDiff$.
 
Alternatively, we have a natural partiality monad $\Lift{(-)}$ on $\wDiff$.
Indeed, for an \wds{} $X$, let us define $\Lift{X}$ as having underlying diffeological space
$\Lift{(|X|,\plots{X})}$ and underlying \wcpo{} $\Lift{(|X|,\leq_X)}$, where we use the previously 
defined partiality monads $\Lift{(-)}$ on $\Diff$ and $\wCpo$.
\begin{proposition}
$\Lift{(-)}$ defines a locally continuous commutative strong monad on $\wDiff$.
Moreover, its Kleisli category $\Lift{\wDiff}$ is equivalent to $\pMap\wDiff$.
\end{proposition}
Again, the important observation is that D-Scott open subsets in $X$ are precisely those subsets with a 
characteristic function $X\to \Lift{\terminal}$ which is an \wds{} morphism.

This category of partial maps has sufficient structure to interpret 
term and type recursion.
\begin{proposition}
    $\wDiff\embed \pMap\wDiff$ forms a bilimit compact expansion.
\end{proposition}
The proof goes via a development of axiomatic 
domain theory \cite{fiore2004axiomatic} in $\wDiff$ that is entirely 
analogous to that given in \cite[Sec. 5]{vakar2019domain} for $\omega$-quasi-Borel spaces.
The only difference are that we work with
\begin{itemize}
    \item the category of $\omega$-dses instead of $\omega$-quasi-Borel spaces;
    \item the domain structure consisting of (the isomorphism closure of) 
    inclusions of D-Scott open subsets with the subspace order and diffeology instead of 
    Borel-Scott open monos
    (in fact, just like in \cite{vakar2019domain}, these are precisely the monos 
    that arise as the pullback of $\return[\terminal]$ along some ``characteristic function''
    $X\to\Lift{\terminal}$);
    \item $T$ equal to $\Lift{(-)}$ as we are not interested in modelling probability.
\end{itemize}
With these minor changes, we can inline Sec. 5 of loc. cit. here to derive bilimit compactness.
We therefore refer the reader to \cite[Sec. 5]{vakar2019domain} for this development.

\begin{corollary}
We obtain a canonical interpretation $\sem{-}$ of the full language of 
Sec. \ref{sec:language2} in $\wDiff$.
\end{corollary}
This corollary follows once we note that $[-,-]$ and the exponential adjunction are
locally continuous as they lift the corresponding locally continuous 
structures in $\wCpo$.
We have extended our previous 
semantics for the limited language, if we forget about the \wcpo{-}structure 
on objects.

We conclude this section by noting that the semantics we have constructed 
is canonical in the following two senses.
First, it is a conservative extension to the full language of Sec. \ref{sec:language2},
respecting all $\beta\eta$-laws, of the canonical semantics of the first-order 
fragment of our language in $\Man$ and $\pMap\Man$. Stronger still: even the semantics of
programs involving recursion and higher-order functions lie in the two canonical 
categories $\Man$ and $\pMap\Man$ for differential geometry, as long as their types are first-order.
Second, this denotational semantics satisfies an adequacy theorem with respect to 
the (completely uncontroversial) operational semantics of our language
(see Appx. \ref{sec:operational-semantics}), showing that our interpretation is a sound means for establishing 
contextual equivalences of the operational semantics.
We hope that these two facts convince the reader of the canonical status of our semantics and 
that they add weight to the correctness theorem of AD with respect to the semantics,
which we prove in the next section.
\section{Logical relations and Correct AD for Recursion}\label{sec:logical-relations2}
\subsection{Preliminaries}
\subsubsection*{Distributive and codistributive laws of monads}
    Suppose we are given two categories each equipped with a monad, $(\catC,S,\return^S,\mu^S)$ and
    $(\catD,T,\return^T,\mu^T)$, together with a functor $F:\catC\to\catD$.
    Then, a natural transformation $\sigma: F;T\to S;F$ is called a \emph{distributive law}
    if it satisfies $\return^T;\sigma=F\return^S$ and $\mu^T_F;\sigma=T\sigma;\sigma_S ;F\mu^S$.
    Similarly, $\tau: S;F\to F;T$ is called a \emph{co-distributive law}
    if it satisfies $F\return^S;\tau=\return^T$ and $F\mu^S;\tau=\tau_S;T\tau;\mu_F^T$.

(Co)distributive laws naturally arise for the following reason \cite{appelgate1965acyclic,mulry1993lifting}.
    Distributive laws $\sigma:F;T\to S;F$ are in 1-1 correspondence with liftings of $F:\catC\to\catD$ 
    to a functor between the Eilenberg-Moore categories: $G:\catC^S\to\catD^T$ such that the left diagram below commutes.
    Co-distributive laws $\tau:S;F\to F;T$ are in 1-1 correspondence with extensions of $F:\catC\to\catD$ 
    to a functor between the Kleisli categories: $H:\catC_S\to\catD_T$ such that the right diagram below commutes.
    Here, we have that $H(A\xto{f}SB)\defeq FA\xto{Ff}F(SB)\xto{\tau}T(FB)$.\vspace{-10pt}
    \begin{figure}[!h]
        \centering
    \begin{tikzcd}
        \catC \arrow[r, "F"]              & \catD             & \catC \arrow[r, "F"] \arrow[d] & \catD \arrow[d]\\
        \catC^S \arrow[r, "G"'] \arrow[u] & \catD^T \arrow[u] &  \catC_S \arrow[r, "H"']        & \catD_T.        
        \end{tikzcd}\vspace{-14pt}\end{figure}

\subsubsection*{Monad liftings}\label{ssec:monad-liftings}
We discussed in Sec. \ref{sec:logical-relations1} how to lift most categorical 
structure to (sub)scones. 
Here, we recall a more systematic way of lifting monads due to \cite{goubault2002logical}.
    Suppose we have monads 
    $(\catC, S, \return^S,\mu^S)$,
    $(\catD,T,\return^T,\mu^T)$, a functor $F:\catC\to\catD$ and a distributive law $\sigma:F;T\to S;F$.
    Then, we obtain a monad $\tilde{T}$ (which lifts $S$
    and $T$) on the scone $\catD/F$ by defining
    \begin{align*}
    \tilde{T}{(D,f,C)}&\defeq (T{D}, T{f};\sigma, S{C})\quad
    \Rreturn[(D,f,C)]\defeq (\return[D]^T,\return[C]^S)\quad
    \Rmu_{(D,f,C)}\defeq (\mu_{D}^T,\mu_{C}^S).
    \end{align*}
    Given an orthogonal factorization system $(\catE,\catM)$ on $\catD$ such that $\catM$
is closed under $T$ and contains $\sigma$, $\tilde{T}$
restricts to a monad on $\catD//F$.
    Suppose that $S$ and $T$ are strong monads with strengths
    $\strength^S$ and $\strength^T$, respectively,
    such that $(\id\times \sigma) ; F(\strength^S)=\strength^T;\sigma$.
    In case $\catC$ and $\catD$ have finite products and $F$ preserves them,
    $\tilde{T}$ gives a strong monad on $\catD/F$.
    If we further assume that $\catM$ is closed under finite products,
    $\tilde{T}$ gives a strong monad on $\catD//F$.
    The strength of $\tilde{T}$ is given by the pair $(\strength^T,\strength^S)$.

\subsection{Logical relations over \wds{}es}
We extend our previous logical relations argument of Sec. \ref{sec:logical-relations1}
to apply to the semantics of recursion in $\wDiff$.
To do so, we extend our definition of the relation to recursive types as
$$
R_{\rec{\tvar}\ty}^U\defeq \set{(\gamma,\gamma') \mid (\gamma;\semroll^{-1},\gamma';\semroll^{-1})\in R_{\subst{\ty}{\sfor{\tvar}{\rec{\tvar}{\ty}}}}^U}.
$$
That is, we would like to define the relation itself using 
type recursion.
Using this definition of the relation, it is not hard to 
prove the fundamental lemma we need for the correctness of 
AD.
The challenge, however, is to show that it is in fact possible 
to define the relation using recursion.
Usually, one shows this using the complicated recipes of \cite{pitts1996relational}.
However, it is not clear that they apply
to the situation we are interested in.
Luckily, however, we are in a situation that is, in some ways,
much simpler than that of \cite{pitts1996relational}: we are 
 considering relations over an $\wCpo$-enriched semantics, rather than the 
non-chain-complete syntax of a language.
As a consequence, it is enough to show that the natural category
of logical relations $\wGl  $ over $\wDiff\times\wDiff$
has a bilimit compact expansion $\wGl  \embed\RLift{\wGl  }$,
where the bilimits in $\RLift{\wGl  }$ lift those in $\Lift{\wDiff}\times\Lift{\wDiff}$.
Then, it automatically follows that our interpretation $\sPair{\sem{-}}{\sem{\Dsyn{-}}}$ 
of our language in $\wDiff\times\wDiff\embed \Lift{\wDiff}\times\Lift{\wDiff}$ lifts to $\wGl  $, showing the 
correctness of AD.

A first attempt might be to define $\wGl  $ as the subscone of the functor 
$\wDiff\times\wDiff\to\Sh(\Opens\RR )$; $(X,Y)\mapsto \plots{X}^{-}\times\plots{Y}^{-}$.
However, the resulting category does not (obviously) give a bilimit compact 
expansion.
Instead, we change the codomain of the functor above with the category $\Sh_{\wCpo}(\Opens\RR )$ of 
internal \wcpos{} in $\Sh(\Opens\RR )$, called  
\emph{\wcpo{-}sheaves} from now, and componentwise \wcont{} natural transformations.
We can characterise these as follows:
\begin{lemma}
An \wcpo{-}sheaf $(|X|,\leq_X)$ in $\Sh(\Opens\RR )$ is the data of a 
sheaf $|X|\in \Sh(\Opens\RR )$ together with a subsheaf $\leq_X$ of $|X|\times |X|$ 
such that $(|X|U, (\leq_X)U)$  forms an \wcpo{} and $|X|(U\subseteq V):|X|V\to|X|U$ is \wcont{}
for each $U\subseteq_{\mathrm{open}} V\subseteq_{\mathrm{open}}\RR $.
\end{lemma}
$\Sh_{\wCpo}(\Opens\RR )$ is $\wCpo$-enriched by using the
order $\alpha\leq \beta\defeq \forall U\in\Opens \RR, x\in U. \alpha_U(x)\leq \beta_U(x)$
on the homsets.
Further,
\begin{proposition}
    $\Sh_{\wCpo}(\Opens\RR )$ is a bicartesian closed category.
\end{proposition}
Indeed, we can show that it is locally presentable, as the category of models of an 
essentially algebraic theory like that of \wcpos{} in any locally presentable category,
e.g. any Grothendieck topos like $\Sh(\Opens\RR )$, is locally presentable \cite[Proposition 1.53]{adamek1994locally}.
Its products are simply taken as the componentwise product in $\wCpo$.
Its coproducts are more complicated, but we shall not need them in generality.
Its exponentials are defined by $|Y^X|C\defeq \Sh_{\wCpo}(\Opens\RR )(\y_C\times X,Y)$, where 
we write $\y$ for the Yoneda embedding, which we equip with the pointwise order.

To construct a subscone of a functor valued in this category, we observe that we have a suitable orthogonal 
factorization system.
First, we note that we have a closure operator $\Close$ on subpresheaves $A$ of $|X|$ for an object 
$X$ of $\Sh_{\wCpo}(\Opens\RR )$: $\Close\, A$ is defined by the smallest chain-complete subsheaf of $|X|$ 
containing $A$.
We can give a predicative construction of $\Close\, A$ using transfinite induction: iterate the closure of 
$A$ in $|X|$ under both lubs of existing $\omega$-chains and amalgamations in $|X|$ of existing matching families in $A$
transfinitely often (potentially creating new $\omega$-chains and matching families at each step).
This iteration defines an ordinal-indexed increasing chain and it must stabilize as it is bounded by the cardinality of $|X|$.
Let us call a subpresheaf $A\subseteq |X|$ \emph{$\Close$-closed} if $\Close\, A =A$
and let us call a morphism $A\to B$ in $\Sh_{\wCpo}(\Opens\RR )$ \emph{$\Close$-dense} if its componentwise image is 
a $\Close$-closed subpresheaf of $|B|$.
\begin{lemma}
Taking $\catM$ to consist of morphisms which are componentwise full monos (i.e. $\Close$-closed monos) 
and $\catE$ to consist of $\Close$-dense morphisms gives an $\wCpo$-enriched orthogonal factorization system on $\Sh_{\wCpo}(\Opens\RR )$.
Further, $\catE$ and $\catM$ are closed under products.
\end{lemma}
The latter statement follows because $\Close$ respects products as a consequence of 
lubs and amalgamations both being taken componentwise in a product.

This factorization system lets us consider the subscone of functors valued in $\Sh_{\wCpo}(\Opens\RR )$.
We have a locally continuous functor $F_{\wGl  }:\wDiff\times\wDiff\to\Sh_{\wCpo}(\Opens\RR )$; $(X,Y)\mapsto (U\mapsto \plots{X}^{U}\times\plots{Y}^{U})$
which preserves finite products.
We write $\wGl  $ for its subscone,
which has triples $(X,Y, R)$ as objects: 
two \wds{}es $X$, $Y$ and a componentwise fully included sub-\wcpo{-}sheaf $R$ of 
$\plots{X}\times\plots{Y}$.
Put simply, we work with pairs $X,Y$ of \wds{es} with relations $R^U\subseteq \plots{X}^U\times\plots{Y}^U$ for each 
$U\subseteq_{\mathrm{open}}\RR $ such that
\begin{itemize}
\item the relations $R^U$ restrict to open subsets and glue along open covers, as before in Sec. \ref{sec:logical-relations2};
\item each $R^U$ is chain-complete: for an \wchain{} $\seq[i\in\NN]{\gamma_i,\gamma'_i}$ with lub $(\gamma,\gamma')$ in $\plots{X}^U\times\plots{Y}^U$ 
such that $\forall i\in\NN. (\gamma_i,\gamma'_i)\in R^U$, we also have $(\gamma,\gamma')\in R^U$.
\end{itemize}
Morphisms $(X,Y,R)\to (X',Y',R')$ in $\wGl  $ are pairs $(X\xto{g}X',Y\xto{h}Y')$ of \wds{} morphisms 
such that, for any $U\in\Opens\RR $, $(\gamma,\gamma')\in R^U$ implies that $(\gamma;g,\gamma';h)\in R'{}^U$.
\begin{proposition}
$\wGl$ is a bicartesian closed category. Further, the copairing
and the exponential adjunction are locally continuous.
\end{proposition}
The first statement follows from the general theory of subsconing from Sec. \ref{sec:logical-relations1}.
The second statement follows as the $\wCpo$-enrichment as well as the coproducts and 
exponential adjunction are lifted from $\wDiff\times\wDiff$, where we have already established this result.
A quick verification shows that the explicit formulae for the logical relation at 
sum and product types given in Sec. \ref{sec:key-ideas} stay valid.
To see the formula for coproducts stays valid, we can note that this 
particular coproduct is straightforward because it is a coproduct of 
subsheaves of sheaves of continuous functions.

Next, we lift the monad $\Lift{(-)}\times\Lift{(-)}$ on $\wDiff\times\wDiff$ to a 
monad $\RLift{(-)}$ on $\wGl  $, using the same definition as before:
$$\RLift{R}^U\defeq \set{\sPair{\gamma}{\gamma'}\in\plots{\Lift{X}}^U\times\plots{\Lift{Y}}^U\mid \gamma^{-1}(X)=\gamma'{}^{-1}(Y)\textnormal{ and }
\sPair{\gamma|_{ \gamma^{-1}(X)}}{\gamma'|_{ \gamma^{-1}(X)}}\in R^{{ \gamma^{-1}(X)}}}.$$
\begin{proposition}
    The strong monad operations of $\Lift{(-)}\times\Lift{(-)}$ lift to $\wGl  $, turning 
$\RLift{(-)}$ into a locally continuous commutative strong monad.
\end{proposition}
Again, we can verify this proposition by hand or appeal to the general results 
of \cite{goubault2002logical}, which we recalled in Sec. \ref{ssec:monad-liftings}.
We discuss the latter option as it will be of further interest.

We have a locally continuous commutative strong 
    monad $\shLift{(-)}$ on $\Sh_{\wCpo}(\Opens\RR)$:\\
    \centerline{$\shLift{F}U\defeq \big(\set{(V,x)\mid V\in\Opens U\textnormal{ and } x\in FV}, ((V,x)\leq(V',x'))\defeq (V\subseteq V'\textnormal{ and }x\leq F(V\subseteq V')(x'))\big)$}\\
    \centerline{$\return[U](x)\defeq (U,x)\quad (V,x)\bind_U \alpha_U\defeq \alpha_V(x).$}
\begin{proposition}
$\Sh_{\wCpo}(\Opens\RR)\embed \shLift{\Sh_{\wCpo}(\Opens\RR)}$ is a bilimit compact expansion.
\end{proposition}
\noindent Again, the proof goes via a development of axiomatic 
domain theory \cite{fiore2004axiomatic} in $\Sh_{\wCpo}(\Opens\RR)$ that is entirely 
analogous to that given in \cite[Sec. 5]{vakar2019domain} for $\omega$-quasi-Borel spaces.
The only difference are that we work with
\begin{itemize}
    \item the category of \wcpos{-}sheaves instead of $\omega$-quasi-Borel spaces;
    \item the domain structure consisting of the monos 
    that arise as the pullback of $\return[\terminal]$ along some ``characteristic function''
    $X\to\shLift{\terminal}$;
    \item $T$ equal to the partiality monad $\shLift{(-)}$ as we are not interested in modelling probability.
\end{itemize}
With these minor changes, we can inline Sec. 5 of loc. cit. here to derive bilimit compactness.

Further, we can observe that we have a distributive law $\sigma:F_{\wGl  };\shLift{(-)}\to \Lift{(-)};F_{\wGl  }$
defined by
\centerline{$(V,(\gamma,\gamma'))\mapsto \left(x\mapsto \left\{\begin{array}{ll}\gamma(x) & x\in V\\
    \bot &\mathrm{else}
\end{array}\right., x\mapsto \left\{\begin{array}{ll}\gamma'(x) & x\in V\\
    \bot &\mathrm{else}
\end{array}\right.\right)$.}\\
Moreover, we can note that $\sigma$ is a full mono and $\Lift{(-)}$ preserves full monos.
As a consequence, we obtain a canonical monad lifting $\RLift{(-)}$ of $\Lift{(-)}$ to the subscone $\wGl$, 
by Sec. \ref{ssec:monad-liftings}. 
Again, the resulting formulas for $\RLift{(R_{\ty}^U)}$ and $R_{\ty\To\ty[2]}^U$ remain the same as those from
Sec. \ref{sec:key-ideas}.

We derive a technical lemma to show that the subscone has a bilimit compact expansion.
\begin{lemma}
    Given locally continuous monads $S$ and $T$ on $\wCpo$-categories $\catC$
    and $\catD$, a locally continuous functor $F:\catC\to\catD$ and a distributive law 
    $\sigma:F;T\to S;F$, giving rise to a monad lifting $\tilde{T}$ to the scone $\catD/F$.
    Suppose that $\catC\embed \catC_S $ and $\catD\embed \catD_T $ are bilimit compact expansions.
    Further assume that 
    \begin{enumerate}
    \item every embedding $\emb{a}$ in $\catC_S$ is a morphism in $\catC$ (factors over $\return$);
    \item $F(\prj{\bot})=\prj{\bot};T{([]_{F\initial})};\sigma$;
    \item $\sigma$ is a split mono with right inverse $\tau$ which is a codistributive law 
    $S;F\to F;T$;
    \end{enumerate}
    And suppose that $\catC\embed \catC_S $ and $\catD\embed \catD_T $ give bilimit compact expansions.
    Then, the Kleisli adjunction $\catD/F\embed (\catD/F)_{\tilde{T}}$ also gives a bilimit compact expansion.

    Further, if we are given a $\wCpo$-enriched factorization system $(\catE,\catM)$ on 
    $\catD$ such that $\sigma\in \catM$ and $\catM$ is closed under $T$, 
    we have seen
    that $\tilde{T}$ restricts to the subscone $\catD//F$.
    Then $\catD//F\embed (\catD//F)_{\tilde{T}}$ also gives a bilimit compact expansion.
\end{lemma}
\noindent In the lemma above, the bilimits in $(\catD/F)_{\tilde{T}}$
lift those in $\catC_S$ and $\catD_T$.
Explicitly, $(\catD/F)_{\tilde{T}}$, the ep-zero object is
$(\initial_{\catD_T},[]_{F\initial},\initial_{\catC_S})$ and
the bilimit of an ep-pair-\wchain{} $\seq[n\in\NN]{(D_n, f_n, C_n), (g_n, h_n)}$ is $((D_\omega, f_\omega, C_\omega), \seq[n\in\NN]{d_n, d'_n})$,
where $(D_\omega, \seq[n\in\NN]{d_n})$ is the bilimit of $\seq[n\in\NN]{D_n,g_n}$ in $\catD_T$,
$(C_\omega,\seq[n\in\NN]{d'_n})$ is the bilimit of $\seq[n\in\NN]{C_n,h_n}$ in $\catC_S$,
and $f_\omega$ is the unique $\catD$-morphism such that $f_\omega;\return^T=
\lub_{n\in\NN}\prj{d_n};T(f_n;F(\emb{{d'_n}}))$.
Remembering that $(\catD//F)$ is a $\wCpo$-enriched-reflective subcategory 
of $\catD/F$ (as $(\catE,\catM)$ is an $\wCpo$-enriched factorization system),
we can construct bilimits in $(\catD//F)_{\tilde{T}}$ by applying the
locally continuous reflector $(\catD/F)\to (\catD//F)$ to
the corresponding bilimits in $(\catD/F)_{\tilde{T}}$.

We observe that in our setting $\sigma$ does have a right inverse $\tau$ which defines a codistributive law:\vspace{-4pt}
\begin{align*}
    &\Lift{(-)};F_{\wGl  }\to F_{\wGl  };\shLift{(-)}\\
&(\alpha,\alpha')\mapsto 
(\Domain{\alpha}\cap \Domain{\alpha'}, (\alpha|_{\Domain{\alpha}\cap \Domain{\alpha'}},\alpha'|_{\Domain{\alpha}\cap \Domain{\alpha'}})).\vspace{-4pt}
\end{align*}
The other conditions of the lemma are also easily seen to be satisfied.
As a consequence, we conclude the following.
\begin{corollary}
$\wGl  \embed \RLift{\wGl  }$ is a bilimit compact expansion, and it lifts the bilimits of 
$\wDiff\embed\Lift{\wDiff}$.
\end{corollary}

We can finally conclude that we can interpret our full language in $\wGl  $.
\begin{corollary}
The interpretation $\sPair{\sem{-}}{\sem{\Dsyn{-}}}$ in $\wDiff\times\wDiff$ 
of the language with type recursion 
lifts to an interpretation $\semgl{-}$ in $\wGl  $.
\end{corollary}
In particular, we can define the relation $R_{\rec{\tvar}\ty}^U$ 
above using type recursion as a part of the interpretation $\semgl{-}$
of our language in $\wGl $.

By repeating the argument of Sec. \ref{ssec:correctness-of-forward-ad} and 
\ref{sec:general-correctness1}, the general correctness theorem \ref{thm:fwd-cor-full} for 
forward AD now extends to the full language with term and type recursion.
We note that it holds for first-order types, which are arbitrary types that do not contain 
any function type constructors. These are all interpreted as objects of $\wDiff$ which are 
isomorphic to manifolds.
In particular, they include algebraic data types such as lists $\rec{\tvar}{\Unit\t+ \reals\t* \tvar}$ and trees
$\rec{\tvar}{\Unit\t+\tvar\t*\reals\t*\tvar}$ holding real numbers.
\section{Related work and concluding remarks}\label{sec:related-work}
This work extends the semantics and correctness proofs of AD
developed in \cite{hsv-fossacs2020,huot2022higher}
to apply to languages with partial features such as recursion.
We model recursion by borrowing techniques developed by \cite{vakar2019domain}
for combining probability and recursion, and we apply them to the 
combination of differentiation and recursion, instead.
\cite{nunes2022automatic,nunes2022logical} present a simplified account 
of the proofs in the present paper.
The algorithms discussed in the present paper can lead 
to efficient reverse AD algorithms, as shown in \cite{smeding2023efficient,smeding2024parallel}.

The first other work on formalization and correctness of AD for 
languages with partiality that we are aware of is \cite{abadi-plotkin2020}.
They show semantic correctness for a define-by-run formulation of reverse AD
on a first-order language with lazy real conditionals and recursion.
In parallel with the present work, \cite{mazza2021automatic} developed a correctness proof for the reverse AD variant of the algorithm discussed in this paper when applied to the PCF language.
Their proofs use operational rather than denotational methods.
An alternative, functional approach to AD that is particularly suitable for reverse AD is CHAD \cite{elliott2018simple,vytiniotis2019differentiable,vakar2021reverse,smeding2024efficient}.
CHAD can be applied to inductive and coinductive types \cite{nunes2023chad} and iteration \cite{2024chaditeration}, but an account of recursion and recursive types is currently still missing from the literature.
We are hopeful that similar proof techniques to those of the present paper should enable a correctness proof of a future CHAD algorithm for recursion and recursive types.

Many popularly used AD systems apply the technique to partially
defined programs built using lazy conditionals, iteration and recursion:
e.g. Stan \cite{carpenter2015stan}, PyTorch \cite{paszke2017automatic},
TensorFlow Eager \cite{agrawal2019tensorflow}, and Lantern \cite{wang2018demystifying}.
This paper makes a contribution towards the mathematical foundations of such 
systems. 
In particular, it demonstrates how to correctly perform forward-mode AD 
on language with partial features.
We plan to use the techniques developed in this paper
to examine the foundations of reverse-mode AD on iterative and recursive language 
features in future work.

\begin{acks}                            
We have benefited from discussing this work with many people, including Mathieu Huot, Ohad Kammar, Sam Staton, and others.
This project has received funding from the European Union’s Horizon 2020 research and innovation
programme under the Marie Skłodowska-Curie grant agreement No. 895827.
\end{acks}


\bibliography{bibliography}

 \clearpage
 \appendix
 \section{Forward AD on coarse-grain CBV}\label{sec:coarse-grain-cbv}
We recall that standard coarse-grain CBV, also known as the $\lambda_C$-calculus,
computational $\lambda$-calculus,
or, plainly, CBV, constructs \cite{moggi1988computational} can be faithfully encoded in
 fine-grain CBV \cite{levy2003modelling,levy2012call}.
This translation $(-)^\dagger$ operates on types and contexts as the identity.
It translates terms $\Gamma\vdash \trm:\ty$ of coarse-grain CBV
into computations $\Gamma\vdash^c \trm^\dagger : \ty$ of fine-grain CBV.
This translation illustrates the main difference between coarse-grain and 
fine-grain CBV: in coarse-grain CBV, values are subset of computations,
while fine-grain CBV is more explicit in keeping values and computations separate.
This makes it slightly cleaner to formulate an equational theory,
denotational semantics, and logical relations arguments.

We list the translation $(-)^\dagger$ below where all newly introduced variables 
are chosen to be fresh.
\[
    \begin{array}{l|l}
        \textnormal{coarse-grain CBV construct } \trm & \textnormal{fine-grain CBV translation } \trm^\dagger\\
        \hline 
        \var & \tReturn \var \\
        \cnst c & \tReturn \cnst c \\
        \tInl \trm & \toin{\var}{\trm^\dagger}{\tReturn\tInl \var}\\
        \tInr \trm & \toin{\var}{\trm^\dagger}{\tReturn\tInr \var}\\
        \tUnit & \tReturn \tUnit\\
        \tPair\trm{\trm[2]} & \toin{\var}{\trm^\dagger}{\toin{\var[2]}{\trm[2]^\dagger}{\tReturn\tPair \var {\var[2]}}}\\
        \fun{\var}{\trm} & \tReturn\fun{\var}{\trm^\dagger}\\
        \tRoll{\trm} & \toin{\trm^\dagger}{\var}{\tRoll{\var}}\\
        \op(\trm_1,\ldots,\trm_n) & \toin{\var_1}{\trm_1^\dagger}{\ldots\toin{\var_n}{\trm_n^\dagger}{\op(\var_1,\ldots,\var_n)}}\\
        \nvMatch{\trm} & \toin{\var}{\trm^\dagger}{\nvMatch{\var}}\\
        \bvMatch{\trm}{\var}{\trm[2]}{\var[2]}{\trm[3]} & \toin{\var[3]}{\trm^\dagger}{\bvMatch{\var[3]}{\var}{\trm[2]^\dagger}{\var[2]}{\trm[3]^\dagger} }\\
        \uMatch{\trm}{\trm[2]} & \toin{\var}{\trm^\dagger}{\uMatch{\var}{\trm[2]^\dagger}}\\
        \pMatch{\trm}{\var}{\var[2]}{\trm[2]} & \toin{\var[3]}{\trm^\dagger}{\pMatch{\var[3]}{\var}{\var[2]}{\trm[2]^\dagger}}\\
        \trm\,\trm[2] & \toin{\var}{\trm^\dagger}{\toin{\var[2]}{\trm[2]^\dagger}{\var\,\var[2]}}\\
        \tItFrom{\trm}{\var}{\trm[2]} & \toin{\var[2]}{\trm[2]^\dagger}{\tItFrom{\trm^\dagger}{\var}{\var[2]}}\\
        \tSign\,\trm & \toin{\var}{\trm^\dagger}{\tSign\,\var}\\
        \rMatch{\trm}{\var}{\trm[2]} & \toin{\var[2]}{\trm^\dagger}{\rMatch{\var[2]}{\var}{\trm[2]^\dagger}}\\
        \rec{\var}{\trm} & \rec{\var}{\trm^\dagger}\\
        \letin{\var}{\trm}{\trm[2]} & \toin{\var}{\trm^\dagger}{\trm[2]^\dagger}.
        \end{array}
\]
This translation induces a semantics $\sem{(-)^\dagger}$ for coarse-grain CBV in 
$\wDiff$, in terms of the semantics $\sem{-}$ of fine-grain CBV.

Moreover, it induces the following forward-mode AD rules for coarse-grain CBV.
The macro $\Dsyn{-}$ on types is as for fine-grain CBV.
On terms, we define a single macro $\Dsyn{-}$ by 
\begin{align*}
\Dsyn{\var}&\defeq \var \\
\Dsyn{\cnst c} & \defeq \tPair{\cnst c}{\cnst 0}\\
\Dsyn{\tInl \trm} & \defeq \tInl \Dsyn \trm\\
\Dsyn{\tInr \trm} & \defeq \tInr \Dsyn \trm\\
\Dsyn{\tUnit} & \defeq \tUnit \\ 
\Dsyn{\tPair\trm {\trm[2]}} & \defeq \tPair{\Dsyn\trm}{\Dsyn{\trm[2]}}\\
\Dsyn{\fun{\var}{\trm}} & \defeq \fun{\var}{\Dsyn\trm}\\
\Dsyn{\tRoll\trm} & \defeq\tRoll{\Dsyn\trm}\\ 
\Dsyn{\op(\trm_1,\ldots,\trm_n)} &\defeq \begin{array}{l}\pMatch{\Dsyn{\trm_1}}{\var_1}{\var'_1}{}\\
    \vdots \\
    \pMatch{\Dsyn{\trm_n}}{\var_n}{\var'_n}
{}\\
\tPair{\op(\var_1,\ldots,\var_n)}{\var'_1*\partial_1\op(\var_1,\ldots,\var_n)+\ldots + \var'_n*\partial_n\op(\var_n,\ldots,\var_n)}\end{array}\\
\Dsyn{\nvMatch{\trm}} &\defeq \nvMatch{\Dsyn{\trm}}\\
\Dsyn{\vMatch{\trm}{\begin{array}{l}\hspace{5pt}\var\To\trm[2]\\\mid\var[2]\To\trm[3]}\end{array}} & \defeq \bvMatch{\Dsyn{\trm}}{\var}{\Dsyn{\trm[2]}}{\var[2]}{\Dsyn{\trm[3]}}\\
\Dsyn{\uMatch{\trm}{\trm[2]}} & \defeq \uMatch{\Dsyn{\trm}}{\Dsyn{\trm[2]}}\\ 
\Dsyn{\pMatch{\trm}{\var}{\var[2]}{\trm[2]}} & \defeq \pMatch{\Dsyn{\trm}}{\var}{\var[2]}{\Dsyn{\trm[2]}}\\
\Dsyn{\trm\, {\trm[2]}} &\defeq \Dsyn{\trm}\,\Dsyn{\trm[2]}\\
\Dsyn{\tItFrom{\trm}{\var}{\trm[2]}} & \defeq \tItFrom{\Dsyn{\trm}}{\var}{\Dsyn{\trm[2]}}\\
\Dsyn{\tSign\trm} & \defeq  \tSign\Dsyn{\trm}\\
\Dsyn{\rMatch{\trm}{\var}{\trm[2]}} & \defeq \rMatch{\Dsyn{\trm}}{\var}{\Dsyn{\trm[2]}}\\
\Dsyn{\rec{\var}{\trm}} & \defeq \rec{\var}{\Dsyn{\trm}}\\
\Dsyn{\letin{\var}{\trm}{\trm[2]}} & \defeq \letin{\var}{\Dsyn{\trm}}{\Dsyn{\trm[2]}}
\end{align*}
where $\var_1:\reals,\ldots,\var_n:\reals\vdash \partial_i\op(\var_1,\ldots,\var_n):\reals$ are 
chosen terms to represent the $i$-th partial derivative of the $n$-ary operation $\op$.
Then, $\Dsyn{\trm}^\dagger = \DsynC{\trm^\dagger}$.

The following correctness theorem for forward AD on coarse-grain CBV is a direct consequence of 
the corresponding correctness theorem for fine-grain CBV.
\begin{theorem*}[Correctness of Fwd AD]
    For any $\var_1:\ty_1,\ldots,\var_n:\ty_n\vdash\trm :\ty[2]$, where $\ty_i,\ty[2]$ are first-order types,
    we have that $\sem{\Dsyn{\trm}^\dagger}(x, v)=(\sem{\trm^\dagger}(x), \Dsemsymbol_{x}\sem{\trm^\dagger}v)$, for all $x$ in the domain of 
    $\sem{\trm^\dagger}$ and $v$ tangent vectors at $x$.
    Moreover, $\sem{\Dsyn{\trm}^\dagger}(x,v)$ is defined iff $\sem{\trm^\dagger}(x)$ is.
    \end{theorem*}
 \section{Operational considerations}
\label{sec:operational-semantics}
The value of our correctness proof of forward AD based on a denotational semantics 
entirely depends on whether the reader believes that the specified semantics is 
in some sense correct.
Recall that our category $\wDiff$ forms a conservative extension 
of the category $\Man$ of manifolds and smooth functions, while $\pMap\wDiff$ 
is a conservative extension of that $\pMap\Man$ of manifolds and smooth partial functions,
completing them to bicartesian closed category with a bilimit compact expansion.
As such, we can interpret a higher-order language with recursive types, 
while the first-order fragment of our language gets its standard interpretation in $\Man$ and $\pMap\Man$, 
possibly the most canonical setting for differential geometry.
Note that even programs which contain higher-order sub-programs get interpreted in $\Man$ and $\pMap\Man$,
as long as their types are first-order.

Still, however, some readers might be interested to understand the connection with the operational 
semantics of our language in order to be convinced that the specified 
denotational semantics is of value.
We detail that correspondence here.

As a fine-grain CBV language, we have an uncontroversal operational semantics:
as a small-step semantics $\trm\leadsto \trm[2]$, we simply use a directed version of the $\beta$-rules of our language, 
supplemented with rules which specify how to evaluate basic operations.
\[
\begin{array}{lll}
    \toin{\var}{\tReturn\val}{\trm}\leadsto\subst{\trm}{\sfor{\var}{\val}}
    &
    \bvMatch{\tInl\val}{\var}{\trm}{\var[2]}{\trm[2]}\leadsto\subst{\trm}{\sfor{\var}{\val}}\\
    \pMatch{\tPair{\val}{\val[2]}}{\var}{\var[2]}{\trm}\leadsto\subst{\trm}{\sfor{\var}{\val},\sfor{\var[2]}{\val[2]}} & 
    \bvMatch{\tInr\val}{\var}{\trm}{\var[2]}{\trm[2]}\leadsto\subst{\trm[2]}{\sfor{\var[2]}{\val}}\\
    (\fun{\var}{\trm})\ \val \leadsto \subst{\trm}{\sfor{\var}{\val}} &
    \rMatch{\tRoll{\val}}{\var}{\trm}\leadsto\subst{\trm}{\sfor{\var}{\val}}\\
    \op(\cnst{c_1},\ldots,\cnst{c_n})\leadsto \tReturn \cnst{\sem{\op}(c_1,\ldots,c_n)} & \textnormal{(for $(c_1,\ldots,c_n)\in \Domain{\sem{\op}}$)}\\
    \tSign\cnst c \leadsto  \tReturn \tInl\tUnit \quad  \tSign\cnst {-c} \leadsto  \tReturn \tInr\tUnit\hspace{-3pt} & \textnormal{(for $c>0$)}.
\end{array}    
\]
Observe that this immediately determines the operational semantics 
for our sugar $\ifelse{\val}{\trm}{\trm[2]}$, $\tItFrom{\trm}{\var}{\val}$,
and $\rec{\var}{\trm}$ as well as for coarse-grain CBV.

From this small-step semantics, we can define a big-step semantics
$\trm\Downarrow \val\defeq \trm\leadsto^* \tReturn \val$,
where we write $\leadsto^*$ for the transitive closure of $\leadsto$.

\emph{Program contexts of type $\ty[2]$
  with a hole $-$ of type $\ctx\vdash\ty$} are terms
$C[\ctx\vdash-:\ty]$ of type $\ty[2]$ with a single variable of type
$\ty$, where this variable $-$ always occurs inside the term in
contexts $\ctx' \geq \ctx$.  Write $C[\trm]$ for the \emph{capturing}
substitution $\subst{C[\ctx \vdash -:{\ty}]}{\sfor{-}{\trm}}$.

Two computations $\Gamma\vdash^c\trm,\trm[2]:\ty$ are in the
\emph{contextual preorder} $\trm\precsim\trm[2]$ when for all program
contexts $C[\ctx \vdash -: {\ty}]$ of type $\reals$, we have that
${C[\trm]\Downarrow}\val$ implies that
${C[\trm[2]]\Downarrow}\val$.
We say that $\trm$ and
$\trm[2]$ are \emph{contextually equivalent}, writing $\trm\approx
\trm[2]$, when $\trm\precsim \trm[2]$ and $\trm[2]\precsim \trm$

We can now state and prove adequacy of our denotational semantics 
with respect to this operational semantics.
\begin{theorem}[Adequacy]
Suppose that two computations $\vdash\trm:\ty$ and $\vdash\trm[2]:\ty$ are comparable in 
our denotational semantics in $\wDiff$ in the sense that 
$\sem{\trm}\leq\sem{\trm[2]}$.
Then, ${\trm}\approx\trm[2]$ in the observational preorder.
In particular, $\sem{\trm}=\sem{\trm[2]}$ implies that 
$\trm\precsim \trm[2]$. 
\end{theorem}
\begin{proof}[Proof (Sketch)]
It is enough to show the corresponding statement for the induced denotational semantics 
$|\sem{-}|$
of our language in $\wCpo$, if we forget about the diffeology.
Indeed, the bicartesian closed structure, as well as the bilimit compact 
expansion structures of $\wDiff$ lift those in $\wCpo$ while also
$\sem{\trm}\leq \sem{\trm[2]}$ iff $|\sem{\trm}|\leq |\sem{\trm[2]}|$.
That is, what we need to do is a standard adequacy
over a semantics in $\wCpo$ of a standard CBV language.
This is precisely the setting where the traditional methods 
of \cite{pitts1996relational} apply.
Indeed, we can simply use a minor extension of the adequacy proof given in  \cite{pitts1996relational}:
we define the logical relation at $\reals$ as $r \trianglelefteq^{\vdash\reals} \cnst{r} $
and we add two more cases to the induction for the fundamental lemma, one for $\tSign$ and one for 
$\op$.
These two steps in the fundamental lemma go through almost tautologically because of our choice 
of denotational semantics precisely matches the operational semantics for these constructs.
Once the fundamental lemma is established, the adequacy theorem again 
follows because the interpretation of values in our semantics remains injective (faithful),
even once we add the type $\reals$ (indeed, values $\underline{c}$ of type $\reals$ 
are in 1-1-correspondence with real numbers $c\in\RR$).
\end{proof}
This shows that our denotational semantics is, in particular,
a sound method for proving contextual equivalences of the operational semantics.
 \section{Characterizing $\wDiff$}
\label{appx:locally-presentable}
We describe three additional categories equivalent to $\wDiff$.
Each approaches the question of how to combine diffeological and \wcpo{}
structures in a different way. We summarise the characterisations:
\ConferenceArxiv{\begin{reptheorem}{7.1}}{\begin{theorem}}
\label{thm:characterisation}
  We have equivalences of categories:
  \(
  \wDiff \equivalent \wE \equivalent \IntwCpo\Diff \equivalent \Mod\wdiffpres \Set
  \).
\ConferenceArxiv{\end{reptheorem}}{\end{theorem}}
We describe these equivalent categories. For our semantics, we only
need these consequences:
\ConferenceArxiv{\begin{repcorollary}{7.2}}{\begin{corollary}}
\label{cor:cocomplete}
  The category of \wdiff{}es, $\wDiff$, is locally
  $\suc\continuum$-presentable, where $\suc\continuum$ is the
  successor cardinal of the continuum. In particular, it has all small
  limits and colimits.
\ConferenceArxiv{\end{repcorollary}}{\end{corollary}}

\subsection{$\wE$: a Domain-Theoretic Completion of Multivariate Calculus}

Diffeological spaces are a quasi-topos completion of the category $\Open$
of open subsets of some $\RR^n$ and smooth functions. Indeed \cite{baez2011convenient}
establish an equivalence
$\DIFF- : \E \equivalent \Diff$, where $\E$ is the full
sub-category of presheaves $[\opposite\Open, \Set]$ consisting of those
functors $F : \opposite\Open \to \Set$ that are:
\begin{itemize}
\item sheaves with respect to open covers; and
\item separated: the functions
$m_F \defeq \seq[r \in U]{F\constantly r} : FU \to \prod_{r \in
  U}F\terminal$ are injective, for all $U\in\Open$.
\end{itemize}
This equivalence is given on objects by mapping $F$ to the diffeological space
$\DIFF F$ whose carrier is $F\terminal$ and whose $U$-plots are
the image $m_F[F U]$ under the injection from the separatedness
condition. Therefore, $m_F : FU \to \prod_{r \in U}F\terminal$ restricts to
a bijection $\xi_F : FU \isomorphic S_{\DIFF F}^U$.
The equivalence is given on morphisms $\alpha : F \to G$
by $\DIFF\alpha \defeq \alpha_{\terminal}$. Given any $\Diff$-morphism
$f : \DIFF F \to \DIFF G$, by setting:
\begin{align*}
\alpha_{U} &:
             FU
             \xto{\xi_F}
             S_{\DIFF F}^U
             \xto{(-);f}
             S_{\DIFF G}^U
             \xto{\xi_G}
             G U
\end{align*}
we obtain the natural transformation $\alpha : F \to G$ for which
$\DIFF\alpha = f$. As an adjoint equivalent to $\DIFF-$, we can choose,
for every diffeological space $X$, the separated sheaf $\invDIFF Xa \definedby X^{a}$
where $a$ is either an object or a morphism, and the (co)unit of this
adjoint equivalence is given by
$\eta_X : \DIFF{\invDIFF X{}} = \invDIFF X\terminal = X^{\terminal}
\xto{\isomorphic} X$.
(See \cite{baez2011convenient} for a more detailed discussion of this equivalence.)

We define a similar category $\wE$ to be the full subcategory of
$[\opposite\Open, \wCpo]$ consisting of those functors
$F : \opposite\Open \to \wCpo$ that are:
\begin{itemize}
\item sheaves; and
\item $\wCpo$-separated: the functions
  $m_F \defeq \seq[r \in U]{F\constantly r} : FU \to \prod_{r \in
    U}F\terminal$ are full monos for all $U\in\Open$.
\end{itemize}
Post-composing with the forgetful functor $\carrier - : \Diff \to \Set$
yields a forgetful functor $\carrier - : \wE \to \E$, as the
underlying function of a full mono is injective.

We equip $\wE$ with an $\wCpo$-category structure by setting the order
componentwise:
\[
  \alpha \leq \beta \qquad\iff\qquad
  \forall U\in \Open. \alpha_U \leq \beta_U
\]

Recall that an \emph{$\wCpo$-equivalence} is an adjoint pair of
locally-continuous functors whose unit and counit are isomorphisms. To
specify an $\wCpo$-equivalence, it suffices to give a fully-faithful
locally-continuous essentially surjective functor in either way,
together with a choice of sources and isomorphisms for the essential
surjectivity.

\ConferenceArxiv{\begin{repproposition}{7.3}}{\begin{proposition}}
  The equivalence ${\wDIFF-} : \wE \equivalent\wDiff$ of
  \ref{thm:characterisation} is $\wCpo$-enriched, and given by:
  \[
    \wDIFF F \definedby
    \triple{\carrier{F\terminal}}{m_F[F\RR]}{\leq_{F\terminal}}
    \qquad\qquad
    \wDIFF {\parent{\alpha : F \to G}} \definedby \alpha_{\terminal}
  \]
  Moreover, the two forgetful functors $\carrier- : \wE \to \E$ and
  $\carrier- : \wDiff \to \Diff$ form a map of adjoints.
\ConferenceArxiv{\end{repproposition}}{\end{proposition}}

\subsection{$\IntwCpo\Diff$: \wcpo s Internal to $\Diff$}
The category $\Diff$ is a Grothendieck quasi-topos, which means it has
a canonical notion of a sub-space: \emph{strong monos}. Sub-spaces let us define
relations/predicates, and interpret a fragment of higher-order logic
formulae as subspaces. In particular, we can interpret \wcpo{}'s
definition \emph{internally} to $\Diff$ and Scott-continuous morphisms
between such \wcpo{}s as a subspace of the $\Diff$-function space, to
form the category $\IntwCpo\Diff$.  We interpret functional operations,
like the $\sup$-operation of a \wcpo{} and the homorphisms of \wcpo s,
as internal functions, rather than as internal functional relations,
as the two notions differ in a non-topos quasi-topos such as $\Diff$.

\subsection{$\Mod{\wdiffpres}\Set$: Models of an Essentially Algebraic Theory}

Both $\wCpo$ and $\Diff$ are locally presentable categories. Therefore,
there are essentially algebraic theories $\wcpopres$ and $\qbspres$
and equivalences $\Mod\wcpopres\Set \equivalent \wCpo$ and
$\Mod\qbspres \Set\equivalent \Diff$ of their categories of
set-theoretic algebras.  We combine these two presentations into a
presentation $\wdiffpres$ for \wdiff{}, given in
the following subsections, by taking their union,
identifying the element sorts and adding a sup operation for \wchain s
of the random elements, and a single axiom stating this sup is
computed pointwise. Local presentability, for example, implies the
existence of all small limits and colimits.

An \wcpo{} or a diffeological space are essentially algebraic, in a precise
sense~\cite[Chapter~3.D]{adamek1994locally}. We will use this algebraic
nature to analyse the diffeological domains and see how the diffeological
structure interacts with the \wcpo{} structure.

\subsubsection{Presentations}
For the following, fix a
regular cardinal $\kappa$. Given a set $\Sorts$ with cardinality
$\cardinality \Sorts < \kappa$, whose elements we call \emph{sorts}, an
\emph{$\Sorts$-sorted ($\kappa$-ary) signature} $\Sig$ is a pair $\Sig
= \pair\Ops\arity$ consisting of a set of \emph{operations} and
$\arity : \Ops \to \Sorts^{<\kappa}\times\Sorts$ assigns to each
operation $\op \in \Ops$ a sequence $\seq[i \in I]{\s_i}$, indexed by
some set $I$ of cardinality $\cardinality I < \kappa$, assigning to
each index $i \in I$ its \emph{argument sort}, together with
another \emph{result sort} $\s$. We write $(\op : \prod_{i \in
  I}{\s_i} \to \s) \in \Sig$ for $\arity(\op) = \pair{\seq[i \in
    I]{\s_i}}\s$.

Given an $\Sorts$-sorted signature $\Sig$, and a $\Sorts$-indexed
sequence of sets $\Var = \seq[\s \in \Sorts]{\Var_s}$ of
\emph{variables} we define the collections of $\Sorts$-sorted terms
$\Term^{\Sig}\Var = \seq[\s \in \Sorts]{\Term^{\Sig}_{\s}\Var}$ over $\Var$
inductively as follows:
\[
  \inferrule{
    ~
  }{
    x \in \Term_{\s}\Var
  }(x \in \Var_{\s})
  \qquad
  \inferrule{
    \text{for all $i \in I$, } t_i \in \Term_{s_i}\Var
  }{
    \op\seq[n \in A]{t_i} \in \Term_{\s}\Var
  }((\op : \prod_{i \in I}{\s_i} \to \s) \in \Sig)
\]
Given a signature $\Sig$ and a sort $\s$, an \emph{equation of sort
  $\s$} is a pair of terms $\pair{t_1}{t_2} \in \Term_{\s}\Var$ over
some set of variables. As each term must involve less than
$\kappa$-many variables, due to $\kappa$'s regularity we may
fix the indexed set of variables $\Var$ to be any specified collection
of sets of cardinality $\kappa$.

\begin{definition}
  An \emph{essentially algebraic presentation} $\Pres$ is a tuple
  $\seq{\Sorts, \tSig, \pSig, \Def, \Eq}$ containing:
  \begin{itemize}
  \item a set $\Sorts$ of \emph{sorts};
  \item two $\Sorts$-sorted signatures with disjoint sets of operations:
    \begin{itemize}
    \item a signature $\tSig$ of \emph{total} operations;
    \item a signature $\pSig$ of \emph{partial} operations;
    \end{itemize}
    we denote their combined signature by $\uSig \defeq
    \pair{\Ops_{\tSig}\union \Ops_{\pSig}}{[\arity_{\tSig},
        \arity_{\pSig}]}$;
  \item for each $(\op : \prod_{i \in I}\s_i \to \s)\in \pSig$, a set
    $\Def(\op)$ of $\tSig$-equations over the variables $\set {x_i :
    \s_i \suchthat i \in I}$ which we call the \emph{assumptions of
    $\op\seq[i \in I]{x_i}$}; and
  \item a set $\Eq$ of $\uSig$-equations which we call the \emph{axioms}.
  \end{itemize}
\end{definition}

The point of this definition is just to introduce the relevant
vocabulary. We will only be considering the following presentation for
posets, then \wcpo s, then diffeological spaces:

\begin{example}[{poset presentation cf.~\cite[Examples~3.35(1),(4)]{adamek1994locally}}]
  The presentation of \emph{posets}, $\pospres$, has two sorts:
  \begin{itemize}
  \item $\elem$, which will be the carrier of the poset; and
  \item $\ineq$, which will describe the poset structure.
  \end{itemize}
  The total operations are:
  \begin{itemize}
  \item $\source : \ineq \to \elem$, assigning to each inequation its lower element;
  \item $\target : \ineq \to \elem$, assigning to each inequation its upper element;
  \item $\refl : \elem \to \ineq$, used to impose reflexivity;
  \end{itemize}
  The partial operations are:
  \begin{itemize}
  \item $\irrelevance : \ineq \times \ineq \to \ineq$, used to impose
    proof-irrelevance on inequations, with $\Def(\irrelevance(e_1, e_2))$:
    \begin{mathpar}
      \source(e_1) = \source (e_2)

      \target(e_1) = \target(e_2)
    \end{mathpar}
  \item $\antisym : \ineq \times \ineq \to \elem$, used to impose anti-symmetry, with\\ $\Def(\antisym(e, \opposite e))$:
    \begin{mathpar}
      \source(e) = \target(\opposite e)

      \target(e) = \source(\opposite e)
    \end{mathpar}
  \item $\trans   : \ineq \times \ineq \to \ineq$, used to impose transitivity, with\\ $\Def(\trans(e_1, e_2))$:
    \begin{mathpar}
      \target(e_1) = \source(e_2)
    \end{mathpar}
  \end{itemize}

    The axioms are:
  \begin{gather*}
    \tag{proof irrelevance}\label{proof irrelevance}
    e_1 = \irrelevance(e_1, e_2) = e_2
    \\
    \tag{reflexivity}\label{reflexivity}
    \source(\refl(x)) = x = \target(\refl(x))
    \\
    \tag{anti-symmetry}\label{anti-symmetry}
    \source(e_1) = \antisym(e_1, e_2) = \source(e_2)
    \\
    \tag{transitivity}\label{transitivity}
    \source(\trans(e_1, e_2)) = \source(e_1)
    \qquad
    \target(\trans(e_1, e_2)) = \target(e_2)
  \end{gather*}
\end{example}

\begin{example}[\wcpo{} presentation]
  In addition to the operations and axioms for posets, the
  presentation $\wcpopres$ of \wcpo{}s includes the following partial
  operations:
  \begin{itemize}
  \item $\lub     : \prod_{n \in \NN}\ineq \to \elem$, used to express lubs of \wchain s, with
    $\Def(\lub_{n \in \NN}e_n)$:
    \begin{mathpar}
      \target(e_n) = \source(e_{n+1}), \text{for each $n \in \NN$}
    \end{mathpar}
  \item for each $k \in \NN$, $\ub_k : \prod_{n \in \NN}\elem \to
    \ineq$, collectively used to impose the lub being an upper-bound, with $\Def(\ub_k\seq[n \in NN]{(e_n)})$:
    \begin{mathpar}
      \target(e_n) = \source(e_{n+1}), \text{for each $n \in \NN$}
    \end{mathpar}
  \item $\least : \elem\times\prod_{n \in \NN} \ineq\times \prod_{n \in \NN}\ineq
    \to \ineq$, used to express that the lub is the least bound, with
    $\Def(\least(x, \seq[n\in\NN]{e_n}, \seq[n \in \NN]{b_n}))$:
    \begin{mathpar}
      \target(e_n) = \source(e_{n+1})

      \target(b_n) = x

      \source(e_n) = \source(b_n), \text{for each $n \in \NN$}
    \end{mathpar}
  \end{itemize}

  The axioms are:
  \begin{gather*}
    \tag{upper bound}\label{upper bound}
    \source(\ub_k\seq[n]{e_n}) = \source(e_k)
    \qquad
    \target(\ub_k\seq[n]{e_n}) = \lub\seq[n]{e_n}
    \\
    \tag{least upper bound}\label{least upper bound}
    \source(\least(x, \seq[n]{e_n}, \seq[n]{b_n})) = \lub\seq[n]{e_n}
    \qquad
    \target(\least(x, \seq[n]{e_n}, \seq[n]{b_n})) = x
  \end{gather*}
\end{example}

\begin{example}[$\diffpres$ presentation]
  The presentation of \emph{diffeological spaces}, $\diffpres$, has
  continuum many sorts:
  \begin{itemize}
  \item $\elem$, which will be the carrier of the diffeological space; and
  \item $\plot_U$, for each $U \in \Open$, which will be the $U$-indexed plots.
  \end{itemize}
  The total operations are:
  \begin{itemize}
  \item $\ev_{ r} : \plot_U \to \elem$, for each $ r \in
    U$, evaluating a plot at $ r \in U$;
  \item $\const : \elem \to \plot_U$ assigning to each element $x$ the
    constantly-$x$ plot; and
  \item $\rearrange_{\phi} : \plot_V \to \plot_U$, for each smooth
    $\phi : U \to V$, precomposing plots with
    $\phi$.
  \end{itemize}
  There are two partial operations:
  \begin{itemize}
  \item $\extensionality : \plot_U\times\plot_U \to \plot_U$, used for
    establishing that a plot is uniquely determined
    extensionally, with $\Def(\extensionality(\alpha, \beta))$ given by
    \[
    \set{\ev_r(\alpha) = \ev_r(\beta) : \elem \suchthat r \in U}
    \]
    and
  \item $\match_{\cover} : \prod_{U \in \cover }\plot_U \to \plot_W$,
    for each open cover $\cover$ of $W$, with
    $\Def(\match_{\cover}\seq[U \in \cover]{f_U})$ given by:
    \[
    \set{\ev_{ r}(f_U) = \ev_{ r}(f_V) \suchthat U, V \in
      \cover,  r \in U\intersect V}
    \]
    used for pasting together a $\cover$-indexed family of compatible
    plots into a case split.
  \end{itemize}

  The axioms are:
  \begin{gather*}
  \tag{extensionality}\label{extensionality}
  \alpha = \extensionality(\alpha, \beta) = \beta
  \\\tag{constantly}\label{const}
  \set{ \ev_{ r}(\const{(x)}) = x \suchthat  r \in U}
  \\\tag{rearrange}\label{rearrange}
  \set{ \ev_{ r}(\rearrange_{\phi}\alpha) = \ev_{\phi( r)}\alpha
    \suchthat  \phi:U \to V\in\Open,  r \in U}
  \\\tag{match}\label{match}
  \set{
    \ev_{ r}\parent{\match_{\cover}\seq[U \in \cover]{\alpha_U}}
    =
    \ev_r(\alpha_U)
    \suchthat 
    \text{$\cover$ open cover of $W$}, U \in \cover, r \in U
    }
  \end{gather*}
\end{example}

We can now present \wdiff{}es:

\begin{example}[\wdiff{} presentation]
  The presentation $\wdiffpres$ of \wdiff{}es extends the presentations
  $\wcpopres$ and $\diffpres$, identifying the $\elem$ sort, with the
  following additional partial operations, for all $U\in\Open$:
  \begin{itemize}
  \item
    $\rlub : \prod_{n \in \NN}\plot_U\times \prod_{n \in \NN, r \in
      U}\ineq \to \plot_U$, used for establishing that the
    plots are closed under lubs w.r.t.~the pointwise order,
    with
    $\Def{\rlub(\seq[n \in \NN]{\alpha_n}, \seq[n \in \NN, r \in
      U]{e_n^r})}$ given by:
    \[
      \set{
        \source(e^r_n) = \ev_r(\alpha_n),
        \target(e^r_n) = \ev_r(\alpha_{n+1}),
        \suchthat n \in \NN, r \in U}
    \]
  \end{itemize}
  The additional axioms are:
  \[
    \tag{pointwise lubs}\label{pointwise lubs}
    \set{
      \ev_r\parent{\rlub\parent{
          \seq[n \in \NN]{\alpha_n},
          \seq[n \in \NN, r \in U]{e_n^r}
        }
      }
      =
      \lub\seq[n \in \NN]{e_n^r}
      \suchthat
      r \in U
    }
  \]
\end{example}

\subsubsection{Algebras}

Every essentially algebraic presentation induces a category of
set-theoretic models, and this category for the \wcpo{} presentation
is equivalent to $\wCpo$. Moreover, we can interpret such
presentations in any category with sufficient structure, namely
countable products and equalisers (i.e., countable limits). We briefly
recount how to do this.

Let $\cat$ be a category with $\lambda$-small limits, with $\lambda$
regular.  As usual, if $\Sig$ is any $\Sorts$-sorted $\lambda$-ary
signature, we define a (multi-sorted) \emph{$\Sig$-algebra} $\Alg =
\pair{\seq[\s \in \Sorts]{\sem{\s}}}{\sem-}$ to consist of an
$\Sorts$-indexed family of objects $\seq[\s \in \Sorts]{\sem\s}$, the
\emph{carrier} of the algebra, and an assignment, to each $\op :
\prod_{i \in I}\s_i \to \s$ in $\Sig$, of a morphism:
\[
\sem\op : \prod_{i \in I}\sem{\s_i} \to \sem\s
\]
Given such an algebra $\Alg$, and an $\Sorts$-indexed set $\Var$ of
variables with $\cardinality{\Var_{\s}} < \lambda$ for each $\s \in
\Sorts$, each term $t$ in $\Term_{\s}\Var$ denotes a morphism:
\[
\sem t_{\s} : \prod_{\s \in \Sorts}\sem{\s}^{\Var_{\s}} \to \sem\s
\]
as follows:
\begin{mathpar}
  \sem{x}_{\s} : \prod_{\s \in \Sorts}\sem{\s}^{\Var_{\s}} \xto{\projection_{\s}}
  \sem{\s}^{\Var_{\s}}
  \xto{\projection_{x}}
  \sem\s

  \sem{\op\seq[i \in I]{t_i}} :
  \prod_{\s \in \Sorts}\sem{\s}^{\Var_{\s}}
  \xto{\seq[i \in I]{\sem{t_i}}}
  \prod_{i \in I}\sem{\s_i}
  \xto{\sem \op}
  \sem{\s}
\end{mathpar}
(When $\cardinality {\Var_{\s}} \geq \lambda$, there are less than
$\lambda$ different variables that actually appear in $t$, and so we
can find a smaller set of sorts and variables for which to define as
above.)

A \emph{$\Sig$-homomorphism} $h : \Alg \to \Alg[2]$ between
$\Sig$-algebras $\Alg$, $\Alg[2]$ is an $\Sorts$-indexed family of
functions $h_{\s} : \Alg\sem{\s} \to \Alg[2]\sem\s$ such that, for
every operation symbol $\op : \prod_{i \in I}\s_i \to \s$ in $\Sig$:
\insertdiagram{01}

We denote the category of $\Sig$-algebras in $\cat$ and their
homomorphisms by $\Mod{\Sig}{\cat}$.


\end{document}